\documentclass[twocolumn,floatfix,superscriptaddress]{revtex4}
\usepackage{amssymb}
\usepackage{graphicx}
\usepackage{lipsum}
\usepackage{dcolumn}
\usepackage{amsmath}
\usepackage{bm}
\usepackage{yhmath}
\usepackage{textcomp}
\usepackage{epstopdf}
\usepackage{color}
\usepackage{ulem}
\usepackage[toc,page,title,titletoc,header]{appendix}
\usepackage[colorlinks,
            linkcolor=blue,
            anchorcolor=blue,
            citecolor=blue
            ]{hyperref}
\usepackage{natbib}
\usepackage{mathrsfs}
\usepackage{soul}

\begin{document}

\title{Non-trivial Interplay of Strong Disorder and Interactions \\
in  Quantum Spin Hall Insulators Doped with Dilute Magnetic Impurities}
\author{Jun-Hui Zheng}\email{jzheng@th.physik.uni-frankfurt.de}
\affiliation{Institut f{\"u}r Theoretische Physik, Goethe-Universit{\"a}t, 60438
Frankfurt/Main, Germany.}
\affiliation{Department of Physics, National Tsing Hua University, Hsinchu 30013,
Taiwan.}
\author{Miguel A. Cazalilla}
\email{miguel.cazalilla@gmail.com}
\affiliation{Department of Physics, National Tsing Hua University, Hsinchu 30013,
Taiwan.}
\affiliation{National Center for Theoretical Sciences (NCTS), Hsinchu 30013, Taiwan.}
\affiliation{Donostia International Physics Center (DIPC), Manuel de Lardizabal, 4. 20018, San Sebastian, Spain.}

\begin{abstract}
We investigate nonperturbatively the effect of a  magnetic dopant impurity on the edge transport of a quantum spin Hall (QSH) insulator. We show that for a strongly coupled magnetic dopant located near the edge of a system, a pair of transmission anti-resonances appear. When the chemical potential is on resonance, interaction effects broaden the anti-resonance width with decreasing temperature, thus suppressing transport for both repulsive and moderately attractive interactions. Consequences for the recently observed QSH insulating phase of the $1$-T$^{\prime}$ of WTe$_2$ are briefly discussed.
\end{abstract}
\pacs{21.XX}
\maketitle

\section{  Introduction}

 Two-dimensional (2D) topological materials like quantum spin Hall insulators (QSHIs) have become a fascinating research topic, with many potential applications~\cite{Hasan2010rmp,exp,review_qshi}. Theoretically, QSHIs are predicted to possess  gapless 
one-dimensional (1D) edge states~\cite{XuMoore2006,review_qshi}.    
Disorder potentials that are invariant under time-reversal symmetry (TRS) cannot cause Anderson localization, which is otherwise ubiquitous in 1D systems. Indeed, it has been shown~\cite{WuBernevigZhang2006,XuMoore2006,Johannesson,
review_qshi} that  for scalar and spin-orbit (SO) disorder potentials, even in the presence of weak electron-electron interactions, the 1D edge channels of QSHIs
exhibit perfect transmission, whose hallmark is a quantized conductance at low temperatures~\cite{Kane20051prl}. On the other hand, strong interactions can break TRS~\cite{WuBernevigZhang2006,XuMoore2006} and lead to complex edge
reconstructions\,\cite{Gefen,Capone}, which jeopardize the perfect conductance quantization.

Experimentally, the QSH effect arising from gapless edge channels has been observed  in HgTe/CdTe and InAs/GaSb/AlSb semiconductor quantum wells (QWs)~\cite{exp}, graphene  submitted to a strong, tilted magnetic field~\cite{young},  
Bi $(111)$ bilayers~\cite{drozdov,yang} and, more recently, in the $1$-T$^{\prime}$  phase of  the transition metal dichalcogenide WTe$_{2}$~\cite{FuLi,Fei2017np,Tang,Pablo}. However, in HgTe/CdTe and InAs/GaSb/AlSb samples, long edge channels  ($\sim1\,\mu$m) in the topological phase exhibit  relatively short mean-free paths, and the conductance deviates from quantization\,\cite{exp,controversy,Du1,Du2}. For the monolayer WTe$_{2}$, the conductance of the devices with longer edges does not exhibit the expected quantized value~ \cite{Fei2017np,Pablo}. Moreover, the interpretation of the observations in InAs/GaSb QWs~\cite{Du1,Du2} has also been questioned  after the discovery of rather similar edge  conduction features in the trivial phase~\cite{controversy},

 Deviations from perfect conductance quantization at low temperatures  arise from backscattering (BS)  in the edge channels. Several BS mechanisms have been discussed using effective 1D models~\cite{Schmidt,Budich,Lezmy,Goldstein,Kainaris,review_qshi}. The latter often involve electron-electron scattering in combination with scalar, spin-orbit coupling and magnetic disorder~\cite{Johannesson,Crepin,TanakaFurusakiMateev2011,Budich,Lezmy,Goldstein,Kainaris,Brataas,Liu2009prl,Kharitonov2017}. Indeed, magnetic impurities break TRS above the Kondo temperature, and therefore they cause BS~\cite{WuBernevigZhang2006,Maciejko,Liu2009prl,TanakaFurusakiMateev2011, Johannesson12}.  Nevertheless, the connection between the effective 1D models of disorder and the 2D aspects of the physics of QSHIs has not yet been fully investigated to the best of our knowledge.  With the exception of a few numerical studies in the non-interacting limit~\cite{Morr,Morr2}, there appears to be no systematic  investigation about the  validity of these 1D models. Indeed, little is known about whether they actually apply in the strong coupling limit where coupling strength to the impurity becomes comparable or larger than the band gap of the QSHI. The latter is an experimentally relevant regime given the small band gaps exhibited by many of the experimentally realized QSHIs. Below, we shall show that the problem of a magnetic dopant impurity problem can be mapped, in the strong coupling limit, to a generalized 1D Fano model~\cite{Fano} describing two resonant levels coupled to an interacting 1D channel. Using  a renormalization group analysis, we show that the transmission coefficient is suppressed at low temperatures for repulsive interactions. Interestingly, when the chemical potential of the edge electrons resonates with one of the in-gap states, we find that the transmission is also suppressed for weak to moderately attractive interactions.
 
The rest of this article is organized as follows: Section~\ref{sec:sol} describes the solution of the scattering problem for a toy model of a single magnetic impurity in the neighborhood of a non-interacting  QSH edge channel. In section~\ref{sec:model}, we construct an effective 1D model to describe this system, which allows us to treat the effect of weak to moderate interactions. In this section, we also discuss the effects not included in  our toy model, such as the Rashba coupling in the band-structure and the non-planar alignment of the magnetic moment. Finally, in section~\ref{sec:sum} we offer the conclusions of this work and provide an outlook for future research directions. The Appendix contains the most technical details of the calculations. Henceforth, we work in units where  $\hbar = 1$.
\section{Model}

In this work, we consider the effect of a magnetic dopant impurity in a QSHI taking into account the electron-electron interactions along the edge. We shall assume a large spin-$S$ magnetic impurity  at temperatures $T$ well above the Kondo temperature  $T_K$ ($T_K$ is  exponentially suppressed for large $S$~\cite{Schrieffer_Kondo}). This allows us to treat the magnetic moment of the dopant classically. For the sake of simplicity, we first solve a model in which the moment lies on the plane perpendicular to the spin-quantization axis of a QSHI, which is described by the Kane-Mele model~\cite{Kane20051prl}. The more general case when the magnetic moment is pointing in an arbitrary direction and the QSHI is described by more realistic extensions  of the Kane-Model model will be discussed in Sect.~\ref{sec:ext}. Once the scattering problem with the  dopant impurity is solved, we  obtain an effective 1D model by fitting the scattering data. The effective model allows us to introduce the electron-electron interactions and treat them non-perturbatively.

 With the above assumptions, the impurity potential is written as follows:
\begin{equation}
\mathcal{V}_{\mathrm{imp}}  =  \lambda_{\mathrm{imp}}
\left(c^{\dag}_{i_0\uparrow} c_{i_0\downarrow}  + \mathrm{h.c.}\right)
= \lambda_{\mathrm{imp}} \: c^{\dag}_{i_0} s^x c_{i_0},
\label{eq:bsi}
\end{equation}
with $c^{\dag}_i  = (c^{\dag}_{i\uparrow},c^{\dag}_{i\downarrow}) $.  As  we will further elaborate below, for $\lambda_{\mathrm{imp}}\gg \Delta$, where $2\Delta$ is the band gap,  two bound states appear within the gap when the impurity  is located deep inside the bulk of the QSHI. 
As the position of impurity is shifted from the bulk to the edge, the bound states hybridize with the edge states inducing a pair of anti-resonances in the transmission coefficient.   Thus, we show that the two-dimensionality arising from the QSHI physics leads to a much richer interplay between interactions and (magnetic) disorder than the one encountered in simple models of structureless impurities in 1D interacting electron systems~\cite{kane1992prl,FisherGlazman,numerics,Natan_BA,Glazman1992,matveev1994prb,Giamarchi,nanotubes}. These results provide the foundation for future  
studies based on more realistic models of the microscopic origins of the absence of quantization in the QSH effect at low temperatures.

 Notice that the model considered here is also drastically different from models based on charge puddles resulting from doping fluctuations~\cite{Goldstein}. Indeed,  the situation envisaged in this work   is more relevant to isolated strongly coupled magnetic moments that are well localized on the lattice scale,  as it is the case of vacancies in 2D materials~\cite{vacancy} or isolated magnetic dopant impurities in general QSHIs.  On the other hand, puddles are described~\cite{Goldstein} as extended quantum dots containing many levels and many electrons, which resonate with the QSH edge states. Furthermore, unlike the study reported below, the authors of Ref.~\cite{Goldstein} neglected Luttinger liquid effects in their treatment of the edge, which may be a good approximation for the HgTe quantum wells due to the large value of the dielectric constant. In the puddle model, backscattering is induced  by the edge electrons dwelling in the quantum dots and undergoing inelastic scattering with other electrons in puddle~\cite{Goldstein}. Thus, in the absence of interactions, the puddle model will not lead to backscattering, whereas the model considered below backscattering is present even in the absence of interactions. 
\section{Solution of scattering problem}\label{sec:sol}
\subsection{Solution of the clean Kane-Mele ribbon}
\begin{figure}[tbp]
\centering
\includegraphics[width=85 mm]{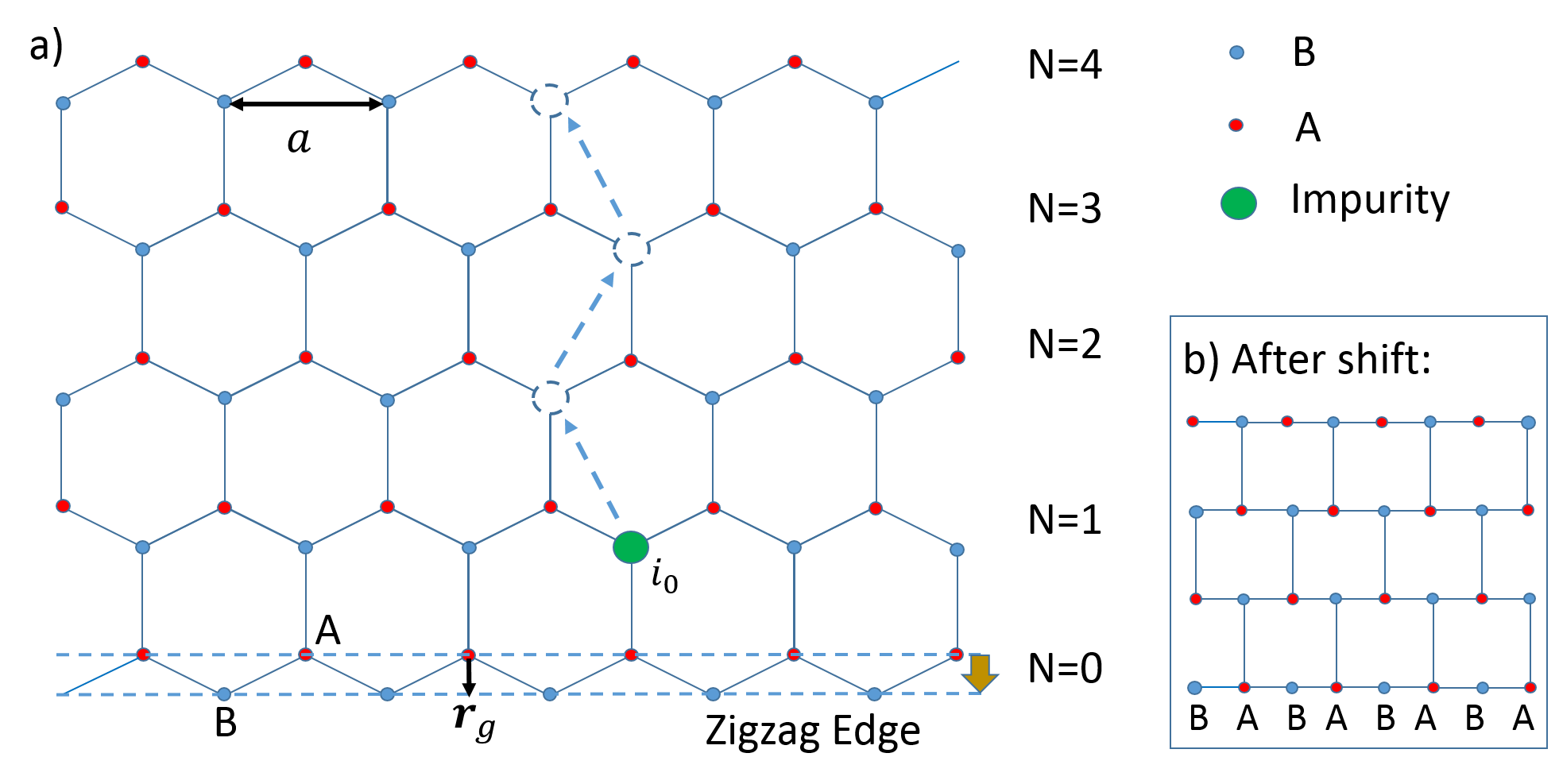}\\
\caption{(Color online) Sketch of (a) the zigzag edge with a single impurity at the edge and (b) the ``brick wall" lattice to which it maps.}
\label{shift}
\end{figure}

 In order to describe the QSHI, we consider the Kane-Mele (KM) model~\cite{Kane20051prl} (cf. Fig.\,\ref{shift}),
\begin{equation}
{H}_{0}=-t\sum_{\langle i,j\rangle }c_{i}^{\dag }c_{j} - i \lambda
_{\mathrm{SO}}\sum_{\langle \langle i,j\rangle \rangle }\nu _{ij}
c_{i}^{\dag }s^{z} c_{j}
\label{Hami-Rx}
\end{equation}
where $\lambda_{\mathrm{SO}}$ describes the {\it intrinsic} SO coupling~\cite{Kane20051prl} as an imaginary next nearest neighbor hopping and $\nu _{ij}=\pm 1$ depends on the electron hoping path; $s^z$ is the electron spin projection on the axis perpendicular to the 2D plane. For the sake of simplicity, we first neglect Rashba SO coupling. This approximation does not qualitatively modify our results, as we discuss in section~\ref{sec:sum}.

In the absence of interactions, the impurity problem is  described by the Hamiltonian:
\begin{equation}
H = H_0 + \mathcal{V}_{\mathrm{imp}}.
\end{equation}
In order to solve this problem, we first obtain an analytical solution of the {\it clean} KM model, Eq.~\eqref{Hami-Rx}, for a zigzag ribbon of  width $L$ (cf. Fig.\,\ref{shift}).
The transmission coefficient of the edge state for the system with an impurity (\ref{eq:bsi}) will be evaluated by solving the Lippmann-Schwinger equation in section~\ref{sec:scat}. 

 In the ribbon geometry, the Bloch wavevector parallel to the edge,  $k_{x}$, is a good quantum number. However,  $k_{y}=-i \partial _{y}$ must be treated as an operator. The wave functions along the $y$-axis obey open boundary conditions~\cite{Zhou20081prl}.  The Hamiltonian \eqref{Hami-Rx} in the Bloch basis can be obtained by using  the Fourier transform,
\begin{equation}\label{gauge}
c_{i \in A}=\sum_{\mathbf{k}}\frac{c_{\mathbf{k}A}}{\sqrt{N_t}}e^{i \mathbf{%
k\cdot }\left( \mathbf{R}_{i}+\mathbf{r}_{g}\right) },\:  c_{i\in B}=\sum_{%
\mathbf{k}}\frac{c_{\mathbf{k}B}}{\sqrt{N_t}}e^{i \mathbf{k\cdot R}_{i }}.
\end{equation}
Here $\mathbf{R}_{i\in  A (B)}$ is the position of $A(B)$ sublattice  sites and $N_t$ is total number of unit cells. Because of the bi-partite structure of the honeycomb lattice, the Fourier transform of $H_0$ is not unique and depends on the relative phase $\mathbf{k}\cdot\mathbf{r}_{g}$. This gauge freedom must be fixed by the boundary conditions (BCs). The appropriate choice for the zigzag edge is
\begin{equation}
\mathbf{r}_{g}=-({a}/{2\sqrt{3}})\mathbf{e}_{y},\end{equation}
 so that the $N$th row of the A sublattice are effectively shifted (See Eq.(\ref{gauge})) to overlap with the $N$th row of the B sublattice (See Fig.\,\ref{shift}). This maps the honeycomb lattice onto the so-called ``brick wall'' lattice and thus  the BCs become
\begin{equation}
\Phi =(\Phi_{B}, \Phi_{A})^T=0 ~~~ {\rm for}  ~~y=\pm L/2.\end{equation}
After identifying the boundary conditions, we  proceed to solve the  1D Schr\"{o}dinger equation:
\begin{equation}
\mathcal{H}_{0}^{s}(\alpha ,\hat\beta ) \Phi _{s}(k_{x},y)=\epsilon \Phi
_{s}(k_{x},y),  \label{Schr-12}
\end{equation}
where we have used the following notation:  $\hat\beta =-i \frac{\sqrt{3}a}{2}\partial _{y}$
and $\mathcal{H}_{0}^{s}=\sum_{i}d_{s}^{i}\sigma ^{i}$, with
\begin{eqnarray} 
d_{s}^{x} &=& -t(2\cos \alpha +\cos \hat\beta ), \notag \\ 
d_{s}^{y} &=& -t\sin \hat\beta, \\ d_{s}^{z} &=& s\lambda _{\mathrm{SO}}(2\sin 2\alpha -4\sin\alpha \cos \hat\beta ), \notag 
\end{eqnarray} 
respectively ($\alpha =k_{x} a /2$).   The Pauli matrices $\sigma^{i}$ ($i = x,y,z$) is in the pseudo-spin space corresponding to the sublattice $\left(B,A\right)$ components.  Furthermore, since $s^z$ is a good quantum number,  $s=\pm 1$. Below, we look for solutions that are combinations of plane waves $e^{ik_y y}$.

We are not interested in finite size effects and therefore take $L\to \infty$.  In this limit, the coupling between the two edges vanishes and we obtain the dispersion for the edge states (see Appendix):
\begin{equation}  
\epsilon _{s}(k_x) =\pm \frac{6s \lambda_{\mathrm{SO}} t \sin (k_x a )}{\sqrt{t^{2}+ \left[ 4\lambda_{\mathrm{SO}}\sin(k_x a/2) \right]^{2}}}, \label{eq:disp}
\end{equation}
where the $+$  ($-$) sign corresponds to the bottom (top) edge at $y = -L/2$ ($y = +L/2$) and $s = \pm 1$. The bands of edge states cross at $k_x  =\tfrac{\pi}{a}$ \cite{Kane20051prl} (for a bearded edge they cross at $k_x  =0$~\cite{Zhang2014Nano}, see appendix). For $k_{x}\approx \tfrac{\pi}{a}$, Eq.~\eqref{eq:disp} agrees with the semi-analytic results of Ref.~\onlinecite{Doh20141arxiv}. For the bottom edge states,  below we use the notation $\vert k_{x},s\rangle$. A plot of the bands~\cite{Kane20051prl} for a wide zigzag ribbon  and the corresponding wavefunctions can be found in the Appendix.
\subsection{Effect of the magnetic impurity}\label{sec:scat}
In order to investigate the effect of the impurity on the electronic transport, we next solve the Lippmann-Schwinger equation (LSE):
\begin{equation}
\left\vert \Psi \right\rangle =\left\vert \Phi \right\rangle +{G} _{0}\left( \epsilon \right) \mathcal{V}_{\text{imp}}\left\vert \Psi \right\rangle,
\end{equation}
where ${G}_{0}\left( \epsilon \right) = (\epsilon + i 0^{+}-H_0)^{-1}$ is the Green's function for Eq.~\eqref{Hami-Rx}.  We  assume the magnetic impurity to be located on the B sublattice at the bottom edge since the wavefunction of edge states on this edge is mostly localized on the  B sublattice (See appendix).  In order to extract the
transmission and reflection coefficients of the edge electrons, we assume the incident electron has a Bloch wave number $k_{x}^{0}$ on the right-moving edge channel, i.e.
$\left\vert \Phi \right\rangle =\left\vert
k_{x}^{0},s=-1\right\rangle$. Therefore, its energy  is $\epsilon_-( k_{x}^{0})$ and its group velocity is $v=\partial_{k_x} \epsilon_-( k_{x}) \vert_{k_{x}=k_{x}^{0}}$. Let us introduce  
\begin{align}
\Phi( s \sigma ,\mathbf{r}) &=\left\langle s,\sigma,\mathbf{r}\right. \left\vert \Phi \right\rangle,\\
\Psi(s \sigma ,\mathbf{r}) &=\left\langle s,\sigma, \mathbf{r} \right. \left\vert \Psi \right\rangle,
\end{align}
where $\sigma=(+,-)$ corresponds to the $(B,A)$ sublattice. Thus,  the asymptotic behavior of the wave function becomes
\begin{eqnarray}
\vert \Psi \rangle &\rightarrow&  (1+\zeta_{t}) \vert \Phi \rangle ~~ {\rm for} ~~x\rightarrow +\infty, \\ 
\vert \Psi \rangle &\rightarrow& \vert \Phi \rangle +\zeta_{r}\vert \tilde\Phi \rangle ~~{\rm for} ~~x\rightarrow -\infty,
\end{eqnarray} 
where $\vert \tilde\Phi \rangle=\left\vert \frac{2\pi}{a} -k_{x}^{0},s=+1\right\rangle$,  and
\begin{align}
\zeta _{t} &=-i \lambda_{\text{imp}}  L_{x} \frac{\Psi (++,\mathbf{r}_{0}) \Phi^{\ast}(-+,\mathbf{r}_{0})}{v},\\
\zeta _{r} &=-i \lambda _{\text{imp}} L_{x} \frac{\Psi ( - +,\mathbf{r}_{0}) \tilde\Phi^{\ast}(++,\mathbf{r}_0)}{v}.
\end{align}
Here $L_x$ is the normalization length of system along the edge  and $\mathbf{r}_0 \propto \textbf{R}_{i_0}$ is the impurity position.

\begin{figure}[tbp]
\centering
\includegraphics[width=3.4 in]{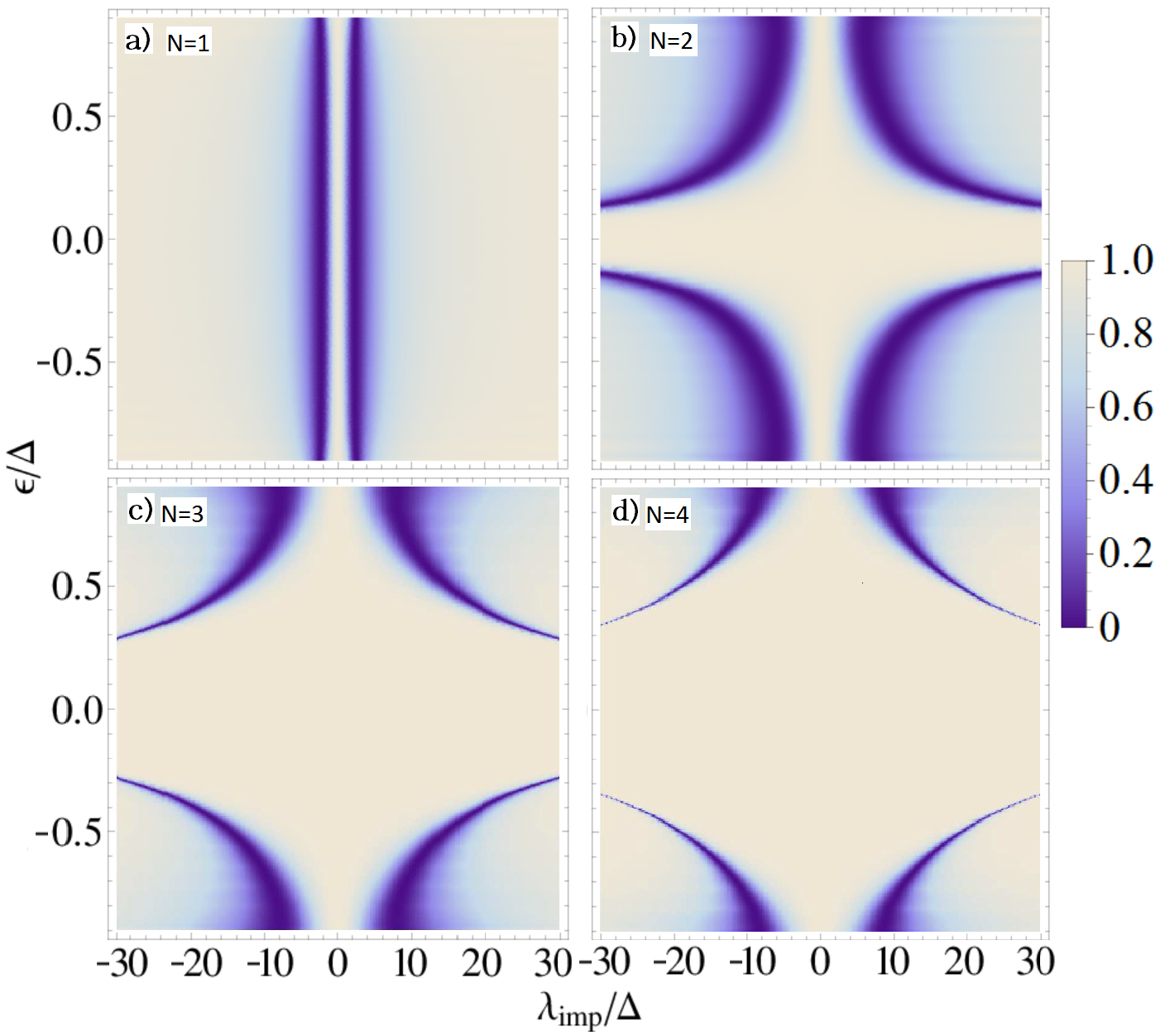}\\
\caption{Transmission coefficient $\mathscr{T}(\epsilon)$ for an impurity on a B sublattice site on (a) the first atomic row (i.e. $N=1$), (b) $N = 2$, (c)  $N=3$, and (d) $N = 4$. The spin-orbit coupling is $\protect\lambda _{\mathrm{SO}}=0.06\,t$.}
\label{transi-2}
\end{figure}

From the above results, the transmission  and the reflection coefficients are obtained from $\zeta_r$
as follows:  
\begin{align}
\mathscr{T}(\epsilon)&=\left\vert 1+\zeta _{t}\right\vert ^{2}\\
 \mathscr{R}(\epsilon) &=\left\vert \zeta _{r}\right\vert ^{2}.
\end{align}
 The energy dependence of the transmission coefficient is shown in Fig.~\ref{transi-2}.  Note that, when the magnetic impurity is located on the first atomic row (i.e. $N=1$), the transmission coefficient is essentially energy independent, which makes it similar to a BS impurity in a purely 1D channel. This behavior arises from  weak coupling between the edge and bulk states via the impurity  (owing to the small weight of the bulk states on the $N=1$ row). This holds true even for relatively large values of  $\lambda_{\mathrm{imp}}$. Thus, scattering is dominated by the 1D edge states. However,  we believe this behavior is not a robust feature but a peculiarity of present KM model. On the other hand, for the second atomic row and beyond (i.e. $N \geq 2$), the weight of the bulk states is larger, and a strong impurity can thus lead to a sizable coupling between bulk and edge states. As a consequence, for  large values of  $\lambda_{\mathrm{imp}}$, a pair of narrow scattering anti-resonances appears within the energy gap. In the neighborhood of the anti-resonances, the transmission coefficient changes very rapidly with energy and, on resonance, it vanishes for large $\lambda_{\mathrm{imp}}$.

In order to understand the emergence of the pair of scattering anti-resonances, we need to consider the poles of the T-matrix,
\begin{equation}
 T(\epsilon) = \left[ \bm{1} - \mathcal{V}_{\mathrm{imp}} G_0(\epsilon)\right]^{-1} \mathcal{V}_{\mathrm{imp}} \label{tmatrix}
\end{equation} 
For a strong impurity potential located within the bulk, of the QSHI, the poles of the T-matrix are obtained from the condition 
\begin{equation}
\mathrm{det}\left[ \bm{1}-\lambda _{\text{imp}} G_{0}^B \left( \mathbf{r}_{0},\mathbf{r}_{0},\epsilon \right) s^x \right]=0
\end{equation}
where  $G_{0}^B$ is the Green's function constructed from bulk states. The latter  is real for $\epsilon$ within the energy gap since the density of states vanishes there and it is odd  in $\epsilon$ (due to the particle-hole symmetry of $H_0$), therefore vanishing at $\epsilon=0$, i.e. the middle of the gap. Thus, $G_{0}^B\left( \mathbf{r}_{0},\mathbf{r}_{0},\epsilon \right) \propto \epsilon$ for small $\epsilon$. Hence, at
large $\lambda_{\mathrm{imp}}$, two bound in-gap states appear at $\epsilon \propto \pm t^2/\lambda_{\mathrm{imp}}$,  corresponding to the two eigenvalues of $s^x$. As the impurity location is shifted towards the edge, the bound states hybridize with the continuum of edge states, leading to the anti-resonances in the transmission coefficient. We will generalize this argument below  in section~\ref{sec:ext} when discussing the effect of extensions to the present toy model. 
\section{1D effective model}\label{sec:model}
\subsection{Non-interacting limit}
After finding a non-perturbative solution to the scattering problem of the edge electrons with a magnetic dopant impurity, in this section we  construct a 1D low-energy effective model that describes a non-interacting edge channel in the presence of magnetic impurity at large $\lambda_{\mathrm{imp}}/\Delta$, where $\Delta=3\sqrt{3}\lambda _{\mathrm{SO}}$ ($2\Delta$ is the bulk band gap). The effective model is valid at energies and temperatures  smaller  than $\Delta$ and therefore only involves the degrees of freedom of the 1D edge and the in-gap states. 

\begin{figure}[t]
\centering
\includegraphics[width=3.2 in]{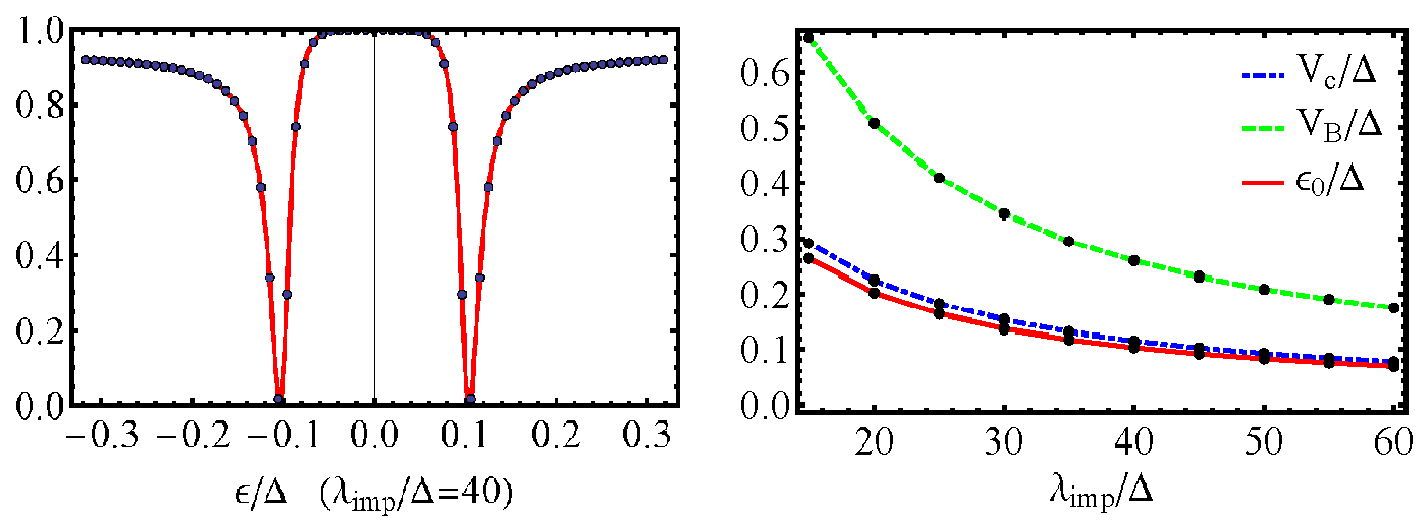} 
\caption{ (Color online)  (Left) Transmission coefficient for an impurity strength $\protect\lambda_{\mathrm{%
imp}}=40\: \Delta$ ($\Delta$ is the band gap). Dots are the transmission coefficient obtained numerically for the Kane-Mele model with a backscatterer at the edge. The red line is the fit to the effective
model (cf. Eq.~\ref{eq:hameff}). (Right) Effective
model   parameters as a function of $\protect\lambda_{\mathrm{imp}}$.}
\label{Transmissions-2}
\end{figure}

  The Hamiltonian of the effective  1D model describing the coupling between the edge electrons and the in-gap states is constrained by the existence of a number of symmetries of  $H = H_0 + \mathcal{V}_{\mathrm{imp}}$. The KM model in the ribbon geometry described by $H_{0}$ (cf. Eq.~\ref{Hami-Rx}), is invariant under TRS ($\mathcal{T}$), spin rotations about the $z$-axis (i.e. $U_{\theta }=\exp (-i\theta s^{z}/2)$, $U^{-1}_{\theta} H_0 U_{\theta} = H_0$), particle-hole symmetry ($\mathcal{C}$), and lattice translations  along the edge direction. The impurity potential, $\mathcal{V}_{\text{imp}}$,  breaks all those symmetries, but  the  composite system described by $H = H_0 + \mathcal{V}_{\text{imp}}$ is invariant under the subgroup span by the combined $U_{\pi}\mathcal{T}$ and $\mathcal{C}\mathcal{T}$ transformations. Therefore, according to the above discussion, the effective model takes the form of a generalized Fano model~\cite{Fano},  describing  two discrete levels coupled to the continuum of edge states. Furthermore, this model is invariant under  $U_{\pi}\mathcal{T}$ and $\mathcal{C}\mathcal{T}$. Since for $|\lambda_{\mathrm{imp}}|\to \infty$ the position of the resonances approaches the center of the band gap at $\epsilon = 0$, we shall focus in the neighborhood of  $k_x = \frac{\pi}{a}$, where linearization of the edge state spectrum, i.e. $\epsilon_{\pm}(k_{x})= \mp v_F k$, is a good approximation. %~\footnote{In general, Eq.~\ref{eq:hamb} is the Hamiltonian describing the clean edge dispersion, e.g. Eq.~\ref{eq:hamb} above}.
 Thus, the effective Hamiltonian can be written as follows:
\begin{align}
{H}_{\mathrm{eff}} &=  H_{B} + H_{+}\left[u,t_{+}\psi(0)\right] + H_{-}\left[d,t_{-}\psi(0)\right],\label{eq:hameff}\\
H_B &= iv_F \int dx\:  \psi^{\dagger }s^{z}
\partial_x\psi + V_{B}a_{0}\psi^{\dagger}(0)s^{x}\psi(0), \label{eq:hamb}\\
H_{\pm}&[f,\chi] = \pm \epsilon_0\: \left( f^{\dag} f -\tfrac{1}{2}\right)
%&\qquad
+ V_{c}a^{1/2}_{0} \left[ f^{\dagger} \chi + \mathrm{h.c.} \right]
%+ \psi^{\dag}(0) t^{\dag}_{\pm} f   \right],
\end{align}
where $\psi^{\dag}(x) = \left(\psi_{L}^\dag(x), \psi^\dag_R(x) \right)$ is the spinor field operator describing the edge states,  $u^{\dag}$ and $d^{\dag}$ are the creation operators of electrons in the bound states with $s^x$ eigenvalue and energy  $s^x = +1, \epsilon = +\epsilon_0$  and $s^x=-1, \epsilon = -\epsilon_0$, respectively, and $t_{\pm} =(\pm 1,1)$; $a_0 = v_F/\Delta$ is a short distance cut-off. In the above model, $V_B$ describes a renormalized backscattering amplitude for the edge electrons, and $V_c$ the tunneling into and out of the bound states. The reflection coefficient for the effective 1D model reads:
\begin{equation}
\mathscr{R}(\epsilon) =  \left\vert \sum_{p= \pm 1} \frac{p}{\frac{i
V_{c}^{2}}{(\epsilon + p \epsilon _{0})\Delta}+( 1- p\frac{i V_{B}}{2\Delta}%
)} \right\vert^{2},
\end{equation}
which accurately fits the results obtained (numerically) for $\mathcal{T}(\epsilon) = 1 - \mathscr{R}(\epsilon)$ from the non-perturbative solution of the scattering problem.  The left panel of Fig.~\ref{Transmissions-2} shows the quality of fit of the transmission coefficient as a function of energy for a magnetic dopant impurity located in the second atomic row (i.e. $N=2$). The behavior of the fitted parameters  $V_{c}$, $V_{B}$, and $\epsilon_0$ as functions of the impurity potential strength $\lambda_{\mathrm{imp}}$ is shown on the right panel. As expected from the above discussion, $\epsilon_0$ decreases as $\lambda_{\mathrm{imp}}\to +\infty$. Note that  $V_c, V_B \ll \Delta$, which is consistent with the assumption that the 1D model, Eq.~\eqref{eq:hameff} describes only the
edge and in-gap states.

\subsection{Interaction effects}\label{sec:inteff}
Finally, we study the effect of electron interactions on the transport properties of the QSHI with a magnetic dopant. Interactions are treated non-perturbatively using the bosonization method~\cite{Giamarchi}. Their characteristic energy scale is $\sim e^2/a_{0}$  (where $e$ is the electron charge), which is assumed to be smaller than the band gap, $2\Delta$.

   In order to apply bosonization to the interacting model, we further project the effective 1D model in Eq.~\eqref{eq:hameff}   onto the subspace of excitations with in the neighborhood of the Fermi energy, $\epsilon_F$. In particular, when $\epsilon_F$ is away from  $\pm \epsilon_0$, the bound states can be integrated out. To leading order, this yields a renormalized backscattering amplitude
\begin{equation}
V^{\prime}_B \simeq V_{B} -  \left[
\frac{V^2_{c}}{\epsilon_0-\epsilon_F} + \frac{V^2_{c}}{\epsilon_0+\epsilon_F}\right]. \label{eq:jbeff}
\end{equation}
and thus the 1D model reduces to the impurity model  in a 1D interacting channel studied by Kane and Fisher~\cite{kane1992prl,FisherGlazman} (cf.  $H_{KF}$ in Eq.~\ref{eq:hamkf} below) with an impurity potential whose backscattering amplitude  $V_B =  V^{\prime}_B$.

On the other hand, on resonance, i.e. for $\epsilon_F \simeq +\epsilon_0$ ($\epsilon_F \simeq -\epsilon_0$), we can integrate out only the non-resonant level at  $\epsilon_F \simeq -\epsilon_0$ ($\epsilon_F \simeq +\epsilon_0$). Assuming (without loss of generality) that  $\epsilon_F \simeq - \epsilon_0$ yields the following low-energy effective model:
\begin{align}
&H^{\prime}_{\mathrm{eff}} = H_{KF} + H_{-}[d,t_{-}\psi(0)] + \left( d^{\dag}d  - \tfrac{1}{2}\right) \notag\\
&\qquad\qquad \times \left[ U_F \: \psi^{\dag}(0)\psi(0) + U_B \: \psi^{\dag}(0)s^x \psi(0) \right], \label{eq:hameff2} \\
&H_{KF}  =  H_B + U \int dx\: \rho_R \rho_L. \label{eq:hamkf}
\end{align}
The interactions between the edge electrons (with amplitude $U$)
and between the edge electrons and the resonant level  (with amplitudes $U_F$ and $U_B$) have been included in the Hamiltonian. We note that integrating out the non-resonant level  at $\epsilon = +\epsilon_0$ renormalizes the amplitude of $V_B - U_B/2$ in $H^{\prime}_{\mathrm{eff}}$  by  an amount $\simeq V^2_c/(\epsilon_F-\epsilon_0)\simeq - V^2_c/2\epsilon_0$. In addition,  forward scattering is also generated but it is dropped since it can be eliminated by a unitary transformation~\cite{kane1992prl,Giamarchi}.

The Hamiltonian $H^{\prime}_{\mathrm{eff}}$ in Eqs.~(\ref{eq:hameff2},\ref{eq:hamkf}) is akin to a model  of a (side-coupled) resonant level  in an interacting 1D channel~\cite{Goldstein2010prlsm,Lerner}. Thus, we apply an analysis similar to the one carried out by Goldstein and Berkovits in Ref.~\cite{Goldstein2010prlsm} to $H^{\prime}_{\mathrm{eff}}$. After bosonizing~\cite{Giamarchi} Eq.~\eqref{eq:hameff2}, we perform a unitary transformation  in order to eliminate the forward interaction term $\propto U_F$ at the expense of renormalizing the scaling dimension ($\Delta_{c}$) of the   operator ($O_c \propto V_c$) describing the tunneling between the 1D edge channel and the resonant level. Thus, 
\begin{equation}
\langle O^{\dag}_c(\tau) O_c(0) \rangle \sim \frac{V^2_c} {\tau^{2\Delta_{T}}},	
\end{equation}
where $\tau$ is the imaginary time and (see Ref.~\cite{Goldstein2010prlsm} and appendix) and
\begin{equation}
\Delta_{T}(K,U_F) = \tfrac{1}{4}\left[K + K^{-1}\left(1-\tfrac{U_F K}{\pi v}\right)^2 \right].\label{eq:delta}
\end{equation}
In this expression
\begin{equation}
K = \sqrt{\frac{2\pi v_F - U}{2\pi v_F + U}}
\end{equation}
is the Luttinger parameter and 
\begin{equation}
v = v_F \sqrt{1 - \left(\frac{U}{2\pi v_F}\right)^2}\label{eq:plasmon}
\end{equation}
the velocity of the edge plasmons~\cite{Giamarchi}. Hence, tunneling into the resonant level becomes relevant in the renormalization-group (RG) sense for
$\Delta_c(K,U_F) < 1$.  There are two different interaction regimes for which this happens: For repulsive interactions (i.e. $K < 1$) and for weak to moderate attraction (i.e. $K \gtrsim 1$).  In the former case,  both tunneling $V_c$ and the BS ($\propto V_B, U_F$) are renormalized  to strong coupling by the  charge-density wave fluctuations dominant in the 1D channel with $K < 1$~\cite{Giamarchi}. At  $T = 0$, transmission through the edge channel is completely suppressed.~\cite{Goldstein2010prlsm,Lerner}

 Interestingly, on resonance the transmission through the edge channel of the QHSI is also suppressed  for moderately attractive interactions i.e. $K \gtrsim 1$. In this regime, backscattering  is {\it na\"ively} irrelevant~\cite{kane1992prl} and therefore  $U_B$ is initially suppressed by the dominant superconducting fluctuations in the edge channel (see below). However, the tunneling amplitude $V_c$  is still a relevant perturbation since $\Delta_c(K, U_F) <1$. Physically, this is because tunneling is a strongly relevant perturbation in 1D, also in the presence of interactions (see e.g. ~\cite{Giamarchi}, chapter 8).

 As the tunneling amplitude renormalizes to strong coupling with decreasing energy scale/temperature, the 2nd order RG flow 
equations (where $y_B \propto U_B$, $y_t\propto V_c$, $\delta_F\propto U_F$, etc. 
are dimensionless couplings,m see appendix~\ref{app:rg} for derivation details):
\begin{align}
\frac{dy_{B}}{d\ln \xi } &=\left( 1-K\right) y_{B}+y_{t}^{2}, \label{eq:rg1} \\
\frac{dy_{t}}{d\ln \xi } &=\left[ 1-K/4- (1-\delta_F)^{2}K^{-1}/4\right]
y_{t}  \notag  \\
&\qquad +y_{t}(y_{B} + v_B), \label{eq:rg2}\\
\frac{d\delta_F}{d\ln \xi} &= 4
(1-\delta_F) y^2_t,  \label{eq:rg3}\\
\frac{d v_B}{d\ln \xi} &= (1-K) v_B.  \label{eq:rg4}
\end{align}
show that this runaway flow of $y_t\propto V_c$  drags along  both the backscattering amplitude $y_B \propto U_B$ and $\delta_F \propto U_F$. This ultimately leads to an effective suppression of the transmission through the edge channel  as the temperature (or the energy scale) is reduced~\cite{Goldstein2010prlsm,Lerner}.

\subsection{Rashba SOC and general magnetic moments}\label{sec:ext}

 The main results obtained using the toy model introduced above can be easily generalized to account for the Rashba spin-orbit coupling in the band structure, i.e. adding to
Eq.~\eqref{Hami-Rx} a term of the form ($\mathbf{d}_{ij}$ is the vector joining the two nearest neighbor sites $i$ and $j$ on the honeycomb lattice):
\begin{equation}
H_{R} =  i \lambda_r \sum_{\langle i, j \rangle} c_i^\dagger (\mathbf{s}\times\mathbf{d}_{ij}) c_j\label{eq:hamrashba}
\end{equation}
and to the case of a more general coupling to the magnetic impurity ($\mathbf{n}$ is a unit vector):
\begin{equation}
\mathcal{\bar{V}}_{\mathrm{imp}} =\lambda_{\mathrm{imp}}  c^{\dag}_{i_0} \left( \mathbf{s} \cdot \mathbf{n}\right) c_{i_0}.\label{eq:genmon}
\end{equation}

In absence of Rashba and for $\mathbf{n}$ perpendicular to the spin-quantization $z$-axis, we can implement rotation along $s_z$ direction to change the magnetic moment in Eq.~\eqref{eq:genmon} to the form  Eq.~\eqref{eq:bsi}, which maps the problem to the toy model studied above.

The presence in the system of a uniform Rashba SOC, Eq.~\ref{eq:genmon}, violates the conservation of the total $s_z$ as well as the particle-hole symmetry of the model. Yet, for weak to moderate Rashba SOC, the topological phase is stable  and exhibits robust helical edge states~\cite{Kane20051prl}.  
In the following, we prove that in the limit $ \lambda _{\text{imp}} \rightarrow \infty $,  a magnetic dopant impurity in the bulk still generates in-gap bound states, which can resonate with the edge states when the impurity is located near the boundary of the insulator.

\begin{figure}[t]
  \centering
  \includegraphics[width=2.6 in]{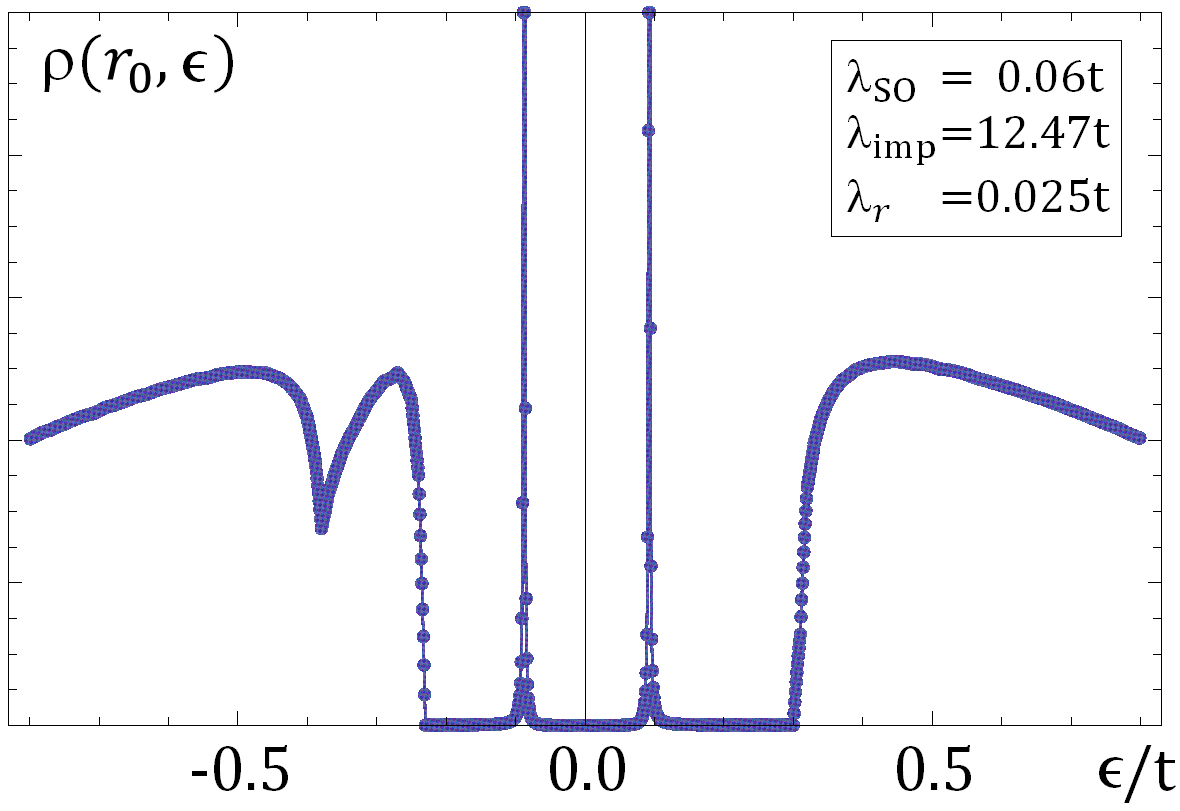}\\
  \caption{Local density of state at the position of a magnetic dopant  impurity located in the bulk of a QSHI insulator described by the Kane-Mele model (see Eqs.\ref{Hami-Rx}   and \ref{eq:hamrashba}) with a strength of the bulk Rashba spin-orbit coupling (SOC) $\lambda_r = 0.06 t$.   The impurity magnetic moment points along the $x$-axis (see Eq.~\ref{eq:bsi}). Notice that the positions of the sharp peaks indicating the existence of impurity-induced in-gap states is not symmetrical with respect to the center of the band gap. This is a consequence of the particle-hole symmetry breaking caused by the Rashba (SOC). }\label{density}
\label{eq:}
\end{figure}

For an arbitrary orientation of the magnetic dopant in 
the bulk of a QSHI,  the positions of bound states are determined by the equation (see Eq. \eqref{tmatrix}):
\begin{equation}	
\mathrm{det}\left[ \bm{1}  - \lambda_{\text{imp}} \left( \mathbf{n}\cdot \mathbf{s} \right) G_{0}^B  \left( \mathbf{r}_{0},\mathbf{r}_{0},\epsilon \right) \right]=0, \label{eq:gencond}
\end{equation}
where  $G_{0}^B(\mathbf{r}_0,\mathbf{r}_0,\epsilon)$ is the local Green's function on the B sublattice, which is  a $2\times 2$ matrix in spin space. However, TRS implies that its off-diagonal
elements vanish~\cite{MeirWingreen,Gurarie}
 $G^B_{0,\uparrow\downarrow} \left( \mathbf{r}_{0},\mathbf{r}_{0},\epsilon \right) = G^B_{0,\downarrow\uparrow} \left( \mathbf{r}_{0},\mathbf{r}_{0},\epsilon \right) =0$
and $G^B_{0,\uparrow\uparrow} \left( \mathbf{r}_{0},\mathbf{r}_{0},\epsilon \right)  = G^B_{0,\downarrow\downarrow} \left( \mathbf{r}_{0},\mathbf{r}_{0},\epsilon \right)$. Hence, $G_{0}^B  \left( \mathbf{r}_{0},\mathbf{r}_{0},\epsilon \right)$ is indeed proportional to the unit matrix, i.e. 
\begin{equation}
G_{0}^B  \left( \mathbf{r}_{0},\mathbf{r}_{0},\epsilon \right)  =  \frac{g_B(\epsilon)}{2}  \: \bm{1},
\end{equation}
where the function $g_B(\epsilon)$ is related to the local
density of states (LDOS) on the B sublattice.   If we apply
a rotation to align the spin quantization axis with the direction of  $\mathbf{n}$, i.e.  $U^{\dag}(\mathbf{n}) \left(\mathbf{n} \cdot \mathbf{s}\right) U(\mathbf{n}) = s^z$, Eq.~\eqref{eq:gencond} yields the following conditions for the existence of in-gap bound states: 
\begin{equation}
 g_B(\epsilon) = \pm 2 \lambda^{-1}_{\mathrm{imp}}\label{eq:ingap}
\end{equation}
The function $g_B(\epsilon)$ 
becomes real for $\epsilon$ within the band gap because the LDOS vanishes there. In addition,  since the LDOS is positive for $\epsilon$ outside the band gap, Kramers-Kronig relationships imply  that 
$g_B(\epsilon)$ must have a zero within the gap, i.e. $ g_B(\epsilon) = z^{-1} (\epsilon - \epsilon_c)$, where $z^{-1}$ is the proportionally constant and $\epsilon_c$ is
an energy within the band gap.
For the KM model, particle-hole symmetry further requires that $\epsilon_c = 0$, which corresponds to the middle of the gap. Rashba SOC breaks particle-hole symmetry and, in general, we expect $\epsilon_c \neq 0$. Hence for sufficiently large $\lambda_{\mathrm{imp}}$, the in-gap states will be located at the energies:
\begin{equation}
\epsilon^{\pm}_{0} = \epsilon_c \pm \frac{2z}{\lambda_{\mathrm{imp}}}. \label{eq:eq}
\end{equation}
However, notice that for $\lambda_{\mathrm{imp}}\sim \Delta$ and/or large particle-hole asymmetry (i.e. $\epsilon_c \sim \Delta$), one or both solutions to Eq.~\eqref{eq:ingap} may not be real. Indeed, this the case when energy of the in-gap states overlaps with the continuum of states in the conduction or valence bands.
However, the above analsys shows that for $\lambda_{\mathrm{imp}}\gg \Delta$, two 
in-gap states will always be present.   The existence of the in-gap bound states can be further  explicitly demonstrated by numerically computing the LDOS of QSHI in the presence of the magnetic dopant impurity.  Fig.~\ref{density} shows the results obtained for the KM with a Rashba SOC of $\lambda_r = 0.06 t$ and $\mathbf{n}$ along the $x$-axis. We have also checked the existence of the in-gap bound state(s) for other choices of $\lambda_r$
and $\mathbf{n}$ (not shown here).

  As the position of the magnetic impurity is shifted towards the edge, the in-gap states hybridize with the topological edge states, which  results in anti-resonances in edge channel transmission. This phenomenon is still described by the generalized Fano model introduced in section~\ref{sec:sol} with different energy values for the energy the in-gap state(s) and the tunneling $V_c$ treated as an energy dependent function. Nevertheless, provided the Fermi level of the 1D edge ($\epsilon_F$) is off resonance, both in-gap states can be integrated out, resulting  in a local backscattering potential, which can be treated as a nonmagnetic impurity in an interacting 1D channel~\cite{Kane20051prl,Glazman1992}. For $\epsilon_F$ on resonance with one of the in-gap state(s), 
the other non-resonant state can be integrated out, giving rise to the similar model to the one studied at the end of section~\ref{sec:inteff}, $H^{\prime}_{\mathrm{eff}}$ (cf. Eq.~\ref{eq:hamkf}, the possible energy dependence of $V_c$ being irrelevant in the RG sense).  A similar argument applies  even when the impurity strength  is not weak or the particle-hole symmetry strong, so that only one bound state exists. An exception to the phenomena described the effective model of Eq.~\eqref{eq:hamkf}  is found when there is a symmetry that prevents the hybridization between the in-gap bound states and the electronic states at the edge. Although this is not generic, it is indeed the case for a dopant whose magnetic moment  $\mathbf{n}$ points along the spin-quantization axis of the KM, Eq.~\eqref{Hami-Rx}. Thus, the total $s^z$ is conserved and the Hilbert space of the problem splits 
into two subspaces labeled by different $s^z$ without
any matrix element connecting them. Thus, conservation of total $s^z$ prevents the existence of backscattering~\cite{TanakaFurusakiMateev2011}.

Therefore, although we have based our calculations in a simplified model of the QSHI and the impurity, the phenomena described above does not depend on the specific microscopic details of the model in the large  $ \lambda _{\text{imp}}$ limit. The emergence of transmission anti-resonances and the interaction induced renormalization of the anti-resonance linewidth~\cite{Goldstein2010prlsm,Lerner} stems from the coupling between the edge states and the impurity-induced in-gap states. This will generically be present as long as the wave functions of the edge states and the states bound by the magnetic impurity overlap. Similar arguments can be applied to magnetic dopants described by more sophisticated models of 
of $\mathbb{Z}_2$ topological insulators.  However,  if $\lambda _{\text{imp}}$ is decreased continuously, the bound states will merge into the continuum of bulk states (together or one by one, depending on the degree of particle-hole asymmetry) and finally the resonances will disappear.

\section{Summary and Outlook}\label{sec:sum}

In summary, we have investigated the transport properties of a quantum spin-Hall insulator in the presence of a strongly coupled magnetic impurity. By obtaining a non-pertubative solution of the scattering problem, we have derived a 1D effective low-energy Hamiltonian describing the system.  In the strong coupling limit, the impurity induces in-gap bound state, which in proximity to the edge state broaden into transmission anti-resonances.  When the chemical potential of the edge electrons is not resonant with any of the in-gap states induced by the magnetic impurity, the system can be effectively mapped to the problem of a nonmagnetic impurity in a Tomonaga-Luttinger liquid~\cite{kane1992prl,matveev1994prb,FisherGlazman} with a renormalized backscattering strength at sufficiently low energy/temperatures (the latter energy scale being set by the separation between the Fermi level and the nearest resonant state).  For strong attractive interactions in the channel, this suppression is absent and the 1D channel becomes increasingly transparent at low $T$.
On the other hand, when the Fermi energy is on resonance, repulsive and weak to moderately attractive interactions lead to temperature-dependent broadening of the transmission anti-resonance, which effectively suppresses the conductance of the edge channel as the temperature $T$ is decreased.  

 For many of the current physical realizations of QSHIs~\cite{exp,Fei2017np,Pablo}, the regime in which $\lambda_{\mathrm{imp}}\gg \Delta$ is not at all unrealistic as the size of the band gap is typically rather small~\cite{exp,review_qshi, Fei2017np,Tang,Pablo}, and its size can be tuned close to the topological transition. In addition, in two-dimensional materials, localized moments can 
appear e.g. from dangling bonds at vacancies~\cite{vacancy}, rather than from magnetic dopants alone. Based on the analysis provided here, 
we believe that the presence of such  localized magnetic defects in proximity to the edge of the recently observed  can induce  significant
backscattering in the newly observed QSHI in the
$1$-T$^{\prime}$ phase of WTe$_2$. The mechanism described here provides additional  backscattering sources to accounts for the experimentally observed~\cite{Fei2017np,Pablo} deviations from conductance quantization at low temperatures. 
Indeed, if the chemical potential of the edge electrons happens to
be at (or near) resonance with in-gap states induced by a magnetic dopant, 
tunneling in/out of the in-gap states will suppress
conductance through the edge channel more effectively than ordinary  backscattering (for comparable strength of the bare backscattering $y_B,v_B$ and tunneling $y_t$ dimensionless couplings, cf. Eqs.~\ref{eq:rg1} to \ref{eq:rg4}). This is because tunneling in/out of the (nearly resonant) in-gap state is a more relevant perturbation than backscattering, as manifested by its smaller scaling dimension (i.e.
typically $\Delta(K, U_F) < K$, cf. Eq.~\ref{eq:delta}), for both repulsive and moderately attractive interactions.  
A more detailed
analysis relevant to this system will be reported in a future publication. Furthermore, in the future we also plan to study extensions to the model studied here beyond the dilute impurity regime (i.e. the multi-impurity case). Another interesting direction is to treat the spin degrees of the magnetic impurity quantum mechanically. This is especially important to describe 
spin-$\tfrac{1}{2}$ impurities below the Kondo temperature. Finally, another interesting research direction, relevant to the study of Majorana bound states, is to the study of the competition of the type of magnetic disorder considered here and
the proximity to a nearby s-wave superconductor~\cite{unpub}. 

We thank L. Glazman, T. Giamarchi, F. Guinea, Y.-H. Ho, C.-L. Huang, and S.-Q. Shen, X.-P. Zhang for useful discussions. M.A.C.  gratefully acknowledges support by the Ministry of Science and Technology (Taiwan) under Contract No. 102- 2112-M-007-024-MY5, and Taiwan's National Center of Theoretical Sciences (NCTS).

\appendix 

\section{Spectrum and wavefunctions}
Here we provide the analytical approach to solve for the spectrum and the wavefunctions of both bulk and edge states for a generalized Kane-Mele (KM) model~\cite{Kane20051prl},
\begin{equation}
\hat{H}_{0}=-t\sum_{\langle i,j\rangle } c_{i}^{\dag }c_{j}- i
\lambda _{\mathrm{SO}}\sum_{\langle \langle i,j\rangle \rangle }\nu _{ij}%
c_{i}^{\dag }s^{z} c_{j}+\lambda _{v}\sum_{i}\xi _{i}
c_{i}^{\dag }c_{i}.  \label{Hami-Rxsm}
\end{equation}
where a staggered potential with $\xi _{i}= + 1$ for $\mathbf{R}_i\in  B$ and $\xi _{i}= - 1$ for $\mathbf{R}_i \in A$ sublattice has been included for generality.  As mentioned in the main text, there is a gauge degree of freedom for the Fourier transformation of $\hat{c}^\dagger_i$ (or $\hat{c}_i$) due to the bi-particle structure of the lattice. The gauge freedom allows to effectively shift the lattice yielding different geometries for the edge. 

 Besides of zigzag edge of interest in the main text, it is also interesting to consider the beard edge in parallel. The   correspond  to two different gauge choices: 1) Zigzag edge: $\mathbf{r}_{g}=-({a}/{2\sqrt{3}})\mathbf{e}_{y}$ and 2) Beard edge: $\mathbf{r}_{g}=({a}/\sqrt{3})\mathbf{e}_{y}$. In our convention, $\sigma^{z}=\left( +,-\right)$ denotes the sublattice pseudo spin components corresponding to the $\left(B,A\right)$ sublattices.

\subsubsection{Spectrum of edge states}

 For the case with zigzag edge, after the Fourier transformation, we obtain
 $\mathcal{H}_{0}^{s}(\alpha,\hat\beta)=\sum_{i}d_{s}^{i}\sigma ^{i}$, we have used the notation where Pauli matrices $\sigma^{i}$ ($i = x,y,z$)  is in the  
pseudo-spin space corresponding to the sublattice $\left(B,A\right)$ components of the single-particle spin wave function and 
\begin{align}
d_{s}^{x}&=-t(2\cos \alpha +\cos \hat\beta ),\\
d_{s}^{y}&=-t\sin \hat\beta,\\  
d_{s}^{z}&=\lambda _{v}+s\lambda _{\mathrm{SO}}(2\sin 2\alpha -4\sin
\alpha \cos \hat\beta ),
\end{align} 
with $\alpha =k_{x} a /2$, and $\hat\beta =-i \frac{\sqrt{3}a}{2}\partial_{y}$.  We set $a=1$ for simplicity. In addition, we treat $\hat\beta$ as an operator and  $\beta$ as its eigenvalues.

Substituting $\Phi _{s}(k_{x},y)=\chi _{s}e^{\kappa y}$  to Eq.\,(\ref{Schr-12}), we get the following secular equation:
\begin{equation}
X f^2 + Y f +Z = 0,
\label{secular-1}
\end{equation}
where the variables 
\begin{align}
f & \equiv \cosh \frac{\sqrt{3}\kappa }{2}, \\ 
 X& = -(4\lambda _{\mathrm{SO}}\sin \frac{k_{x}}{2})^{2}, \\
 Y &= 8s\lambda _{\mathrm{SO}}\sin \frac{k_{x}}{2}\left( \lambda _{v}+2s\lambda _{\mathrm{SO}}\sin k_{x}\right) -4t^{2}\cos \frac{k_{x}}{2}, \\
Z&=\epsilon^{2}-t^{2}-4t^{2}(\cos \frac{k_{x}}{2})^{2}-\left( \lambda _{v}+2s\lambda _{\mathrm{SO}}\sin k_{x}\right) ^{2}.
\end{align} 
Hence, we obtain the following two roots:
\begin{equation}
 f_{1,2}=({-Y\pm \sqrt{Y^{2}-4XZ}})/{2X},
\end{equation}
Thus,  there are four roots for $\kappa $,  corresponding to $\pm \kappa _{1,2}$ with $\kappa _{1,2}=\frac{2}{\sqrt{3}}\cosh ^{-1}f_{1,2}$. For the edge states, we have that $\mathrm{Re}\: \kappa_{1,2} \neq 0$. Thus, we use the convention that $\mathrm{Re}\: \kappa _{1,2}>0$ for the function $\kappa _{1,2}=\frac{2}{\sqrt{3}}\cosh^{-1}f_{1,2}$.

Note that only two linearly independent wavefunctions satisfy the open boundary conditions corresponding the zigzag edge, namely $\Phi _{s}(k_{x},\pm L/2)=0$ for each value of $\epsilon$. They are 
\begin{align}
&g_{c}^{1}(k_{x},y)-g_{c}^{2}(k_{x},y), \\
&g_{s}^{1}(k_{x},y)-g_{s}^{2}(k_{x},y), 
\end{align}
where~\cite{Zhou20081prl} 
\begin{align}
g_{c}^{i}(k_{x},y)&=\frac{\cosh (\kappa _{i}y)}{\cosh (\kappa _{i}L/2)}, \\ 
g_{s}^{i}(k_{x},y) &=\frac{\sinh (\kappa _{i}y)}{\sinh (\kappa _{i}L/2)}.
\end{align}
The eigenfunctions can be expressed as the linear combination of the above wavefunctions. By introducing a 
$2 \times 2$ matrix of coefficients $\mathcal{L} = \left[ l_{ij}\right]$, the eigenfunctions 
can be written as follows:
\begin{equation}
\Phi _{s}(k_{x},y)=\mathcal{L}\left[
\begin{array}{c}
g_{c}^{1}(k_{x},y)-g_{c}^{2}(k_{x},y) \\
g_{s}^{1}(k_{x},y)-g_{s}^{2}(k_{x},y)%
\end{array}%
\right].  \label{wave-1}
\end{equation}
Substituting this function into Eq.~\eqref{Schr-12}, and using that $g_{c,s}^{i}$ are linearly independent, we obtain the following conditions relating the column vectors of the matrix $\mathcal{L}$:
\begin{align}  \label{relationsforL}
\mathbf{L}_{2} &=\tanh (\kappa _{1}L/2)M_{1}\mathbf{L}_{1},\\
\mathbf{L}_{2} &= \tanh (\kappa _{2}L/2)M_{2}\mathbf{L}_{1}, \\
\mathbf{L}_{1} &=\frac{1}{\tanh (\kappa _{1}L/2)}M_{1}\mathbf{L}_{2},\\
\mathbf{L}_{1} &= \frac{1}{\tanh (\kappa _{2}L/2)}M_{2}\mathbf{L}_{2},
\end{align}
where 
\begin{align}
\mathbf{L}_{1} &=(l_{11},l_{21})^{T}, \\
\mathbf{L}_{2} &=(l_{12},l_{22})^{T}, \\
M_{i} &=\sigma ^{y}\left\{\left(-2t\cos \alpha -t\cos \beta_{i}\right)\sigma ^{x} \right. \notag \\
&\qquad \left. 
 + \left(\lambda _{v}+2s\lambda_{\mathrm{SO}}\sin 2\alpha \right. \right.  \notag\\
&\left. \left. \quad -4s\lambda_{\mathrm{SO}}\sin \alpha \cos \beta_{i}\right)\sigma ^{z}  -\epsilon\right\}/(t\sin \beta _{i}),\\ 
\beta _{i}&=-i \frac{\sqrt{3}}{2}\kappa _{i},
\end{align}
%\iffalse  
respectively. Note that, in the  above derivation, we have used the fact $\cos \hat\beta g_{c,s}^{i}(k_{x},y)=\cos \beta _{i}g_{c,s}^{i}(k_{x},y)$, $\sin\hat\beta g_{c}^{i}(k_{x},y)=\sin \beta _{i}\tanh (\kappa_{i}L/2)g_{s}^{i}(k_{x},y)$ and $\sin \hat\beta g_{s}^{i}(k_{x},y)=\frac{\sin \beta _{i}}{\tanh (\kappa _{i}L/2)}g_{c}^{i}(k_{x},y)$.
%\fi

The combinations of equations in the same column in Eq.\eqref{relationsforL} give the secular equation \eqref{secular-1}, which  relates $\kappa _{i}$ and spectrum $\epsilon $. The other two independent equations are obtained by combining diagonal terms in Eq.\eqref{relationsforL}, which yields:
\begin{equation}
\mathbf{L}_{2}=TM_{1}M_{2}\mathbf{L}_{2}=\frac{1}{T}M_{2}M_{1}\mathbf{L}_{2},
\label{relation-1}
\end{equation}%
where 
\begin{equation}
T=\frac{\tanh (\kappa _{1}L/2)}{\tanh (\kappa _{2}L/2)}.
\end{equation}
 Expressing $\kappa _{i}$ as functions of $\epsilon$, this equation is exactly the
constraint for spectrum $\epsilon$. In the following, we will solve this equation. Eq.\eqref{relation-1} implies that
\begin{equation}
M_{t}\mathbf{L}_{2}=0,
\end{equation} 
where $M_{t}\equiv TM_{1}M_{2}-\frac{1}{T}M_{2}M_{1}$. To have a nontrivial solution for $\mathbf{L}_{2}$, the condition $\det M_{t}=0$ is required, which gives 
\begin{equation}
(T+\frac{1}{T})^{2}={4D_{0}^{2}}/({{D_{0}^{2}-D_{x}^{2}-D_{y}^{2}-D_{z}^{2}}}),\end{equation} 
where
\iffalse Explicitly,
%
\begin{equation}
M_{t}=\frac{-T\left[ D_{0}+D_{x}\sigma _{x}+D_{y}\sigma _{y}+D_{z}\sigma _{z}%
\right] +\frac{1}{T}\left[ D_{0}-D_{x}\sigma _{x}-D_{y}\sigma
_{y}-D_{z}\sigma _{z}\right] }{t^{2}\sin \beta _{1}\sin \beta _{2}},
\end{equation}
%
where
\fi
%
\begin{align}
D_{x}&=t(\cos \beta _{1}-\cos \beta _{2})\epsilon,\\
 D_{y}&=i t(\lambda_{v}+6s\lambda _{\mathrm{SO}}\sin 2\alpha )(\cos \beta _{1}-\cos \beta _{2}),\\
 D_{z}&=4s\lambda _{\mathrm{SO}}\sin \alpha (\cos \beta _{1}-\cos \beta
_{2})\epsilon,
\end{align}
and $D_{0}=t^{2}(2\cos \alpha +\cos \beta _{1})(2\cos
\alpha +\cos \beta _{2})+(\lambda _{v}+2s\lambda _{\mathrm{SO}}\sin 2\alpha
-4s\lambda _{\mathrm{SO}}\sin \alpha \cos \beta_{1})(\lambda _{v} 
 +2s\lambda
_{\mathrm{SO}}\sin 2\alpha -4s\lambda _{\mathrm{SO}}\sin \alpha \cos \beta
_{2})-\epsilon ^{2}$.
For $L\rightarrow\infty$,  $T=1$ because $\mathrm{Re}\: \kappa _{1,2}>0$. Thus, it becomes 
\begin{equation}
D_{x}^{2}+D_{y}^{2}+D_{z}^{2}=0,
\end{equation}
which gives the dispersion
\begin{equation}\label{speccc}
\epsilon _{s}^{\pm }=\pm \frac{t(\lambda _{v}+6s\lambda _{\mathrm{SO}}\sin
2\alpha )}{\sqrt{t^{2}+(4\lambda _{\mathrm{SO}}\sin \alpha )^{2}}}.
\end{equation}
%.

\begin{figure*}[tbp]
\centering
\includegraphics[width=12 cm]{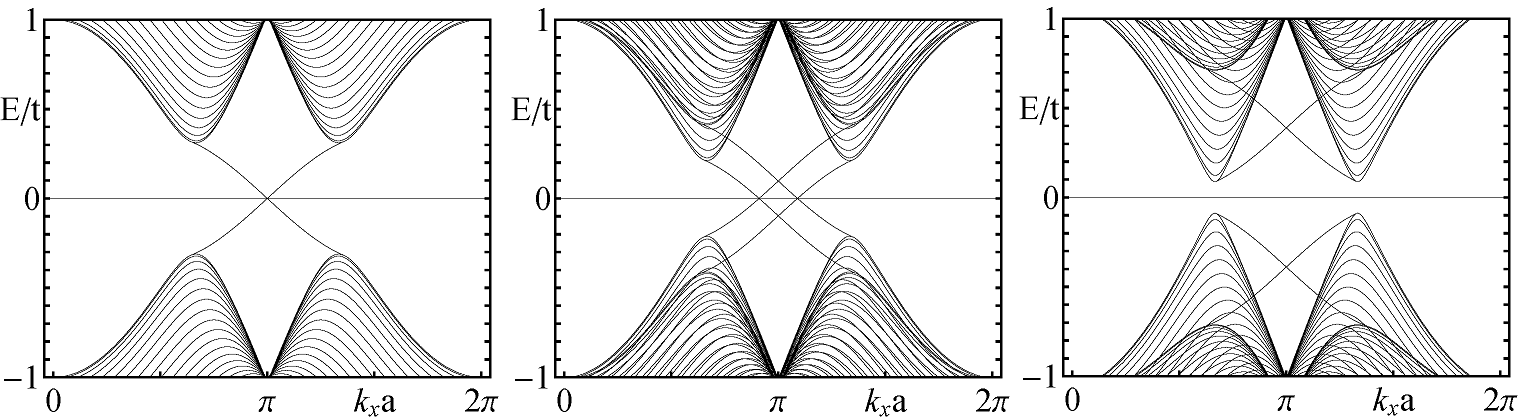}
\caption{Band structure of a zigzag ribbon described by Eq.~\eqref{Hami-Rxsm}. Left panel: $\protect\lambda %
_{\mathrm{SO}}=0.06t$ and $\protect\lambda _{v}=0$; Central panel: $\protect%
\lambda _{\mathrm{SO}}=0.06t$ and $\protect\lambda _{v}=0.1t<3\protect\sqrt{3%
}\protect\lambda _{\mathrm{SO}}$; Right panel: $\protect\lambda _{\mathrm{SO}%
}=0.06t$ and $\protect\lambda _{v}=0.4t>3\protect\sqrt{3}\protect\lambda _{%
\mathrm{SO}}$.}
\label{spectrum-Z}
\end{figure*}

Note that Eq.\eqref{relation-1} gives an additional constraint for the spectra. From $M_{t}\mathbf{L}_{2}=0$ and $T=1$, we obtain 
\begin{equation}(D_{x}\sigma_{x}+D_{y}\sigma _{y}+D_{z}\sigma _{z})\mathbf{L}_{2}=0.
\end{equation}
 Combining it with $\mathbf{L}_{2}=M_{1}M_{2}\mathbf{L}_{2}$ (Eq.(\ref{relation-1})), we obtain the following the constraint:
\begin{equation}  \label{conditions}
-D_{0}=t^{2}\sin \beta _{1}\sin \beta _{2}.
\end{equation}
The second constraint is $\mathrm{Re}\: \kappa _{1,2}>0$. These constraints restrict a region $k_{x}\in \left[ \Lambda _{s}^{-},\Lambda _{s}^{+}\right] $, where edge states exist. We show the resulted spectra in 
Fig.~\ref{spectrum-Z}.

\subsubsection{Wavefunctions for a semi-infinite system}

To investigate the wavefunctions at one of edges only, it is helpful to shift the coordinate origin so that the QSHI occupies the upper half plane ($0\leq y\leq L$ with $L\to\infty$). For a semi-infinite system, the wavefunction satisfying the boundary condition $\Phi _{sk_{x}}(0)=\Phi _{sk_{x}}(L\rightarrow \infty)=0$ has a much simpler form:
\begin{equation}
\Phi _{sk_{x}}(y)=C_{s}(k_{x})\chi _{s}(k_{x})(e^{-\kappa
_{s,1}y}-e^{-\kappa _{s,2}y}).  \label{wave-semi}
\end{equation}
Thus, what is left is to determine the $2\times 1$ matrix $\chi _{s}(k_{x})$ and the normalization factor $C_{s}(k_{x})$. For each $k_x$, we have obtain the
spectra $\epsilon_{s}$ and the wave number $\kappa_{s,i}=\frac{2}{\sqrt{3}} \cosh ^{-1}f_{s,i}$ with $i=1,2$ in the last section. Substituting
Eq.~\eqref{wave-semi} into the Schr\"odinger Eq.(\ref{Schr-12}), we obtain
\begin{equation}
\chi _{s}(k_{x}) =\Bigg[
\begin{array}{c}
-\left[ \mathcal{H}_{0}^{s}\right] _{12}/\left\{ \left[ \mathcal{H}_{0}^{s}%
\right] _{11}-\epsilon_{s}\right\} \\
1%
\end{array}
\Bigg] \equiv \Bigg[
\begin{array}{c}
\chi_{s,1} \\
1%
\end{array}%
\Bigg].
\end{equation}
Explicitly,
\begin{widetext}
\begin{equation}
\chi_{s,1} =\frac{2t\cos \frac{k_{x}}{2}+t\exp (%
{\sqrt{3}\kappa _{s,1}}/{2})} {\lambda _{v}+2 s \lambda _{\mathrm{SO}} [\sin
k_{x} - 2\sin ({k_{x}}/{2}) \cosh ({\sqrt{3} \kappa _{s,1}} / {2}
)]-\epsilon_s}
\end{equation}
\end{widetext}
Recall that the above wavefunctions only make sense
when evaluated on the discrete set of points of the honeycomb  lattice:
\begin{equation}
\psi _{sk_{x}}(n)=C_{s}(k_{x})\chi _{s}(k_{x})(e^{-\kappa
_{s,1}ny_{0}}-e^{-\kappa _{s,2}ny_{0}}),
\end{equation}
where $y_{0}=\sqrt{3}a/2$. The normalization factor is
\begin{equation}
C_{s}(k_{x})=(1+ \left\vert \chi _{s,1}\right\vert 
^{2})^{-1/2}C_{s}^{0}(k_{x})
\end{equation}
where
\begin{align}
C_{s}^{0}(k_{x}) &=\left[ \Upsilon\left(2\: \mathrm{Re}\: \kappa _{s,1}\right) +\Upsilon \left( 2\: \mathrm{Re}\: \kappa_{s,2}\right)
\right. \notag\\
&\left.
-\Upsilon\left(\kappa _{s,1}^{\ast}+\kappa _{s,2}\right) 
-\Upsilon\left( \kappa_{s,1}+ \kappa_{s,2}^{\ast}\right) \right]^{-1/2}
\end{align}
and  $\Upsilon(k) \equiv {1}/[{%
1-\exp ( -k y_{0}) }]$. Upon denoting $\Phi _{sk_{x},\sigma }(\mathbf{r})$ as the
$\sigma$ components of $\Phi _{sk_{x}}(\mathbf{r})$, we find $|\Phi
_{sk_{x},+ }(\mathbf{r})|^2 \gg |\Phi _{sk_{x},-}(\mathbf{r})|^2$ for the
case $\lambda _{v}=0$ and $\lambda %
_{\mathrm{SO}} \ll t$, which suggesting the bottom edge states `prefer' B-sublattice.

\subsubsection{Wavefunctions for Bulk States}
For the bulk states with periodic boundary conditions, crystal momentum $
\mathbf{{k}=(}k_{x},k_{y}\mathbf{)}$ is treated as good quantum number
in both the $x$-direction and $y$-direction. Thus,
upon setting $\kappa =ik_{y}$ in Eq.~\eqref{secular-1}
(with $\beta =\frac{\sqrt{3}k_{y}}{2}$), we obtain the (bulk)
dispersion:
\begin{widetext}
\begin{equation}
E_{s\eta }\left( \mathbf{k}\right) = E_{s\eta }\left( k_x, k_y\right) =\eta \sqrt{t^{2}+4t^{2}\cos \alpha \cos
\beta +4t^{2}\cos ^{2}\alpha +[\lambda _{v}+2s\lambda _{\mathrm{SO}}(\sin
2\alpha -2\sin \alpha \cos \beta )]^{2}},
\end{equation}
\end{widetext}
where $s,\eta =\pm 1$.

However, for  open boundary conditions and in
the limit $L\rightarrow \infty $, the spectrum of bulk state is not modified
from the above form because the boundary effects become negligible in the thermodynamic limit.  On other hand,  wavefunctions are modified and become different from Bloch waves because of the scattering with the boundary.
Thus,  from the secular equation \eqref{secular-1}, for each $\kappa _{1}=i k_{y}$($k_y$ is real) and thus $f_{1}\equiv \cos \frac{\sqrt{3}k_{y}}{2}$, we can
find another root, $f_{2}=-\frac{Y}{X}-f_{1}$. In total four
different roots for $\kappa$ exist, i.e. $\pm \kappa _{1,2}$ with $\kappa _{1,2}=\frac{2}{\sqrt{3}}\cosh ^{-1}f_{1,2}$, corresponding to a same energy $\epsilon$. Note that $f_{1}$ and thus $f_{2}$ are real. Thus there are two different cases: 1) $\left\vert f_{2}\right\vert >1$, the plane wave decays at the edge; and 2) $\left\vert f_{2}\right\vert \leq 1$, different modes interference with each other:

{\textbf{Case 1:}} For $\left\vert f_{2}\right\vert >1$, we have $\kappa _{2}= \frac{2}{\sqrt{3}}\cosh ^{-1}f_{2}$ with $\mathrm{Re}\: \kappa _{2}>0$. Thus
the full solutions of the secular equation~\eqref{secular-1} for $\kappa$
are $\pm i k_{y}$ and $\pm \kappa_{2}$. The mode $\sim e^{\kappa _{2} y}$
diverges for $y\rightarrow\infty$, so it will not emerge and there are
only there modes left: $e^{\pm \kappa _{1} y}$ and $e^{-\kappa _{2} y}$.
After using the boundary condition $\Phi _{s\eta ,\mathbf{k}}(y=0)=0$, only
two linear independent wavefunctions are left. The general wavefunction has
the following form:
\begin{equation}
\Phi _{s\eta ,\mathbf{k}}(y)=\frac{C_{s\eta }(\mathbf{k})}{\sqrt{N_{y}}}%
\mathcal{L}\Bigg[
\begin{array}{c}
\exp (i k_{y}y)-\exp (-\kappa _{2}y) \\
\exp (-i k_{y}y)-\exp (-\kappa _{2}y)%
\end{array}%
\Bigg] ,  \label{wave-b}
\end{equation}
where $\mathcal{L=}\left[ l_{ij}\right] _{2\times 2}$ is a $2\times 2$
matrix, and $C_{s\eta }(\mathbf{k})$ is the normalization constant.
Obviously, such a kind of wavefunction is a combination of extended state
and local state, which decays at the edge.

Now we need to calculate out the matrix $\mathcal{L}$. Substituting Eq.(\ref%
{wave-b}) into Schr\"{o}dinger equation~\eqref{Schr-12}, and using the fact
that $\exp (\pm ik_{y}y)$ and $\exp (-\kappa _{2}y)$ are linear independent,
we obtain the following results: 
\begin{align}
\mathbf{L}_{1}=c_{1}\Bigg[%
\begin{array}{c}
l_{1} \\
1%
\end{array}%
\Bigg], &~~~~~
\mathbf{L}_{2}=c_{2}\Bigg[%
\begin{array}{c}
l_{1}^{\ast } \\
1%
\end{array}%
\Bigg],\\
 \mathbf{L}_{1}+\mathbf{L}_{2}&=\Bigg[%
\begin{array}{c}
l_{2} \\
1%
\end{array}%
\Bigg],
\end{align} 
where  
\begin{align}
\mathbf{L}_{1} &=(l_{11},l_{21})^{T},\\ 
\mathbf{L}%
_{2}&=(l_{12},l_{22})^{T},\\
l_{i}&=-\frac{\left[
\mathcal{H}_{0}^{s}(\alpha ,\beta _{i})\right] _{12}}{ \left[ \mathcal{%
H}_{0}^{s}(\alpha ,\beta _{i})\right] _{11}-E_{s\eta }\left( \mathbf{k}%
\right) }
\end{align}
and $c_{1}$, $c_{2}$ are constants, $\beta _{1}=\frac{\sqrt{3}}{2}k_{y}$, $\beta _{2}=i\frac{%
\sqrt{3}}{2}\kappa _{2}$. 
Solving these equations, we find $c_{1}=%
\frac{l_{2}-l_{1}^{\ast }}{l_{1}-l_{1}^{\ast }}$ and $c_{2}=\frac{%
-l_{2}+l_{1}}{l_{1}-l_{1}^{\ast }}$.

The next step is to calculate the normalization coefficient $C_{s\eta }(\mathbf{k})$. For large $L$ limit, $\exp (-\kappa _{2}y)$ does not influence
normalization. Using the orthogonality of $\exp (\pm ik_{y}y)$, we obtain
\begin{equation}
C_{s\eta }(\mathbf{k})=\frac{1}{\sqrt{\left\vert c_{1}\right\vert
^{2}+\left\vert c_{2}\right\vert ^{2}}\sqrt{\left\vert l_{1}\right\vert
^{2}+1}}.
\end{equation}%
As a result, in real space, we have
\begin{equation}
\Phi _{s\eta ,\mathbf{k}}(n)=\frac{C_{s\eta }(\mathbf{k})}{\sqrt{N_{y}}}%
\mathcal{L}\Bigg[%
\begin{array}{c}
\exp (ik_{y}ny_{0})-\exp (-\kappa _{2}ny_{0}) \\
\exp (-ik_{y}ny_{0})-\exp (-\kappa _{2}ny_{0})%
\end{array}%
\Bigg],
\end{equation}
{\textbf{Case 2:}} $\ $For $\left\vert f_{2}\right\vert \leq 1$, we
have $\kappa _{2}=\frac{2}{\sqrt{3}}\cosh ^{-1}f_{2}=ik_{y}^{\prime }$ with $%
k_{y}^{\prime }\geq 0$. The full solutions of the secular equation
\eqref{secular-1} for $\kappa $ are $\pm ik_{y}$ and $\pm ik_{y}^{\prime }$.
The boundary conditions $\Phi _{s\eta ,\mathbf{k}}(y=0)=0$ require these
four running waves inference with each other, and  thus there are only
three linear independent wavefunctions. Following the method used in previous
case, we can construct the eigenfunctions by combining the three
wavefunctions. However, we shall proceed in a different way here.
Similar to the previous case, there is one eigenfunction,
\begin{equation}
\left\vert 1\right\rangle =\frac{1}{\sqrt{N_{y}}}C_{s\eta }(\mathbf{k})%
\mathcal{L}\Bigg[%
\begin{array}{c}
\exp (ik_{y}y)-\exp (-ik_{y}^{\prime }y) \\
\exp (-ik_{y}y)-\exp (-ik_{y}^{\prime }y)%
\end{array}%
\Bigg],
\end{equation}
where $\mathcal{L}$ is same as the one in Eq.~\eqref{wave-b} except for the replacement of $\kappa _{2}$ with $ik_{y}^{\prime}$ and thus  the normalization becomes:
\begin{equation}
C_{s\eta }(\mathbf{k})=\frac{1}{\sqrt{%
(|c_{1}|^{2}+|c_{2}|^{2})(|l_{1}|^{2}+1)+(|l_{2}|^{2}+1)}}.
\end{equation}
The second eigenstate $\left\vert 2\right\rangle $ can be obtained by the
replacements: $k_{y}\to k_{y}^{\prime }$ (which implies that
$l_{2}\to l_{1}^{\ast }$). We denote the corresponding
parameters as $\mathbf{L}_{1}^{\prime }$, $\mathbf{L}_{2}^{\prime}$, $%
c_{1}^{\prime }$, $c_{2}^{\prime }$ and $C_{s\eta }(\mathbf{k}^{\prime })$. Note that these two eigenstates are not orthogonal.

In the following, we construct an orthogonal and symmetric basis by means of
\begin{equation}
\left\vert +\right\rangle =\left\vert 1\right\rangle +\vartheta \left\vert
2\right\rangle ,~~~~\left\vert -\right\rangle =\left\vert 2\right\rangle
+\vartheta \left\vert 1\right\rangle.
\end{equation}
Using the orthogonality condition together with $\left\langle 1\right.
\left\vert 1\right\rangle =\left\langle 2\right. \left\vert 2\right\rangle
=1 $, we obtain
\begin{equation}
\left\vert \vartheta \right\vert ^{2}=1,~~~\text{Re}\:\vartheta =-\text{Re}\:
\left\langle 1\right. \left\vert 2\right\rangle ,
\end{equation}
where $\left\langle 1\right. \left\vert 2\right\rangle =C_{s\eta }(\mathbf{k}%
)C_{s\eta }(\mathbf{k}^{\prime })[-c_{2}^{\ast
}(|l_{1}|^{2}+1)-c_{2}^{\prime }(|l_{2}|^{2}+1)]$. We use the convention
that $\mathrm{Im}\: \vartheta =\sqrt{1-\left( \mathrm{Re}\:\vartheta \right) ^{2}}\geq 0$, and finally, we obtain the orthonormalized wavefunctions
\begin{eqnarray}
\Phi _{s\eta \mathbf{k}}(n) &=& \frac{1}{\sqrt{2+2\mathrm{Re} d \left[ \vartheta
\left\langle 1\right. \left\vert 2\right\rangle \right] }}\left\vert
+\right\rangle, \\
\Phi _{s\eta \mathbf{k}^{\prime }}(n) &=& \frac{1}{\sqrt{2+2%
\text{Re}\left[ \vartheta \left\langle 2\right. \left\vert 1\right\rangle %
\right] }}\left\vert -\right\rangle.
\end{eqnarray}

\begin{figure*}[btp]
\begin{center}
\includegraphics[width=12 cm]{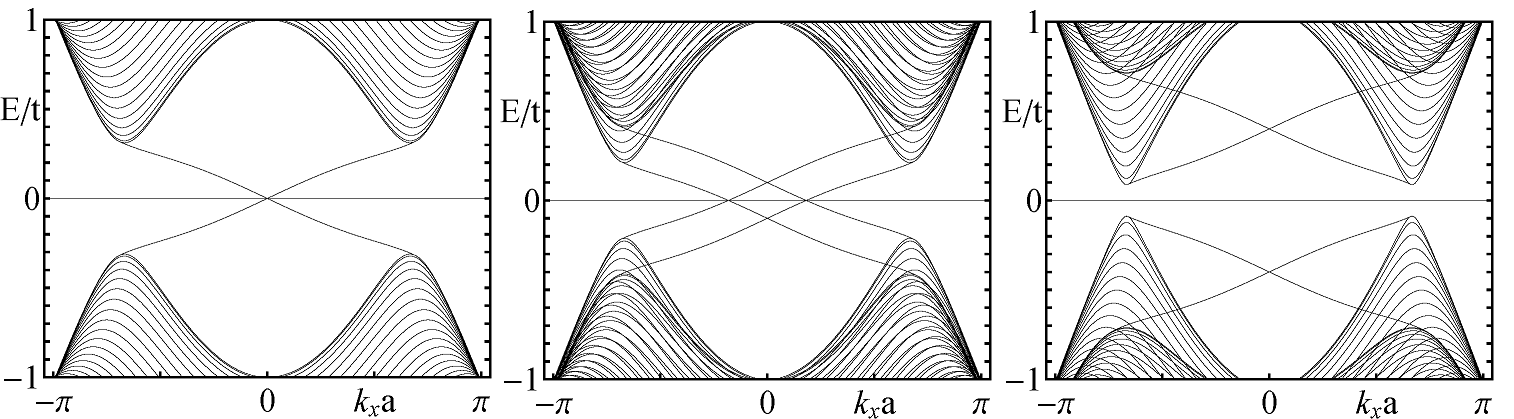} 
\end{center}
\caption{Band structure of a bearded-edge ribbon described by Eq.~\ref{Hami-Rxsm}. Left panel: $\protect\lambda %
_{\mathrm{SO}}=0.06t$ and $\protect\lambda _{v}=0$; Central panel: $\protect%
\lambda _{\mathrm{SO}}=0.06t$ and $\protect\lambda _{v}=0.1t<3\protect\sqrt{3%
}\protect\lambda _{\mathrm{SO}}$; Right panel: $\protect\lambda _{\mathrm{SO}%
}=0.06t$ and $\protect\lambda _{v}=0.4t>3\protect\sqrt{3}\protect\lambda _{%
\mathrm{SO}}$.}
\label{spectrum-B}
\end{figure*}
\section{Green's function for the Kane-Mele model in a semi-infinite system}
So far, we have obtained the eigenvalues and the complete set of eigenfunctions for the model of Eq.~\ref{Hami-Rxsm}. Hence, the Green's function can be expressed
in terms of them:
\begin{equation}
\hat{G}_{0}\left( \epsilon \right) =\sum_{\mathbf{k},s\eta }\frac{\left\vert
\mathbf{k},s,\eta \right\rangle \left\langle \mathbf{k},s,\eta \right\vert }{%
\epsilon +i0^{+}-E_{s\eta }(\mathbf{k})}+\sum_{k_{x},s}\frac{\left\vert
k_{x},s\right\rangle \left\langle k_{x},s\right\vert }{\epsilon
+i0^{+}-\epsilon _{s}\left( k_{x}\right) },
\end{equation}
where $\left\vert \mathbf{k},s,\eta \right\rangle =\Phi _{s\eta \mathbf{k}}$.
In the real space, 
\begin{equation}G_{0,\sigma \sigma ^{\prime }}^{s}\left( \mathbf{r},%
\mathbf{r}^{\prime },\epsilon \right) =\left\langle \mathbf{r,}s,\sigma
\right\vert \hat{G}_{0}\left( \epsilon \right) \left\vert \mathbf{r}^{\prime
}\mathbf{,}s,\sigma ^{\prime }\right\rangle,
\end{equation}  where $\sigma =\pm 1$
represents the different components of $\sigma ^{z}$. Thus, using $%
\left\langle \mathbf{r,}s,\sigma \right. \left\vert \mathbf{k},s,\eta
\right\rangle =\Phi _{s\eta \mathbf{k,}\sigma }(\mathbf{r})$ and $%
\left\langle \mathbf{r,}s,\sigma \right. \left\vert k_{x},s\right\rangle
=\Phi _{sk_{x},\sigma }(\mathbf{r})$, where $\Phi _{s\eta \mathbf{k,}\sigma
}(\mathbf{r})$ and $\Phi _{sk_{x},\sigma }(\mathbf{r})$ are the $\sigma $
components of $\Phi _{s\eta \mathbf{k}}(\mathbf{r})$ and $\Phi _{sk_{x}}(%
\mathbf{r})$ respectively, we have
\begin{align}
G_{0,\sigma \sigma ^{\prime }}^{s}\left( \mathbf{r},\mathbf{r}^{\prime
},\epsilon \right) & =\sum_{\mathbf{k},\eta }\frac{\Phi _{s\eta \mathbf{k,}%
\sigma }(\mathbf{r})\Phi _{s\eta \mathbf{k,}\sigma ^{\prime }}^{\ast }(%
\mathbf{r}^{\prime })}{\epsilon +i0^{+}-E_{s\eta }(\mathbf{k})} \notag \\
&~~~~ +\sum_{k_{x}}%
\frac{\Phi _{sk_{x},\sigma }(\mathbf{r})\Phi _{sk_{x},\sigma ^{\prime
}}^{\ast }(\mathbf{r}^{\prime })}{\epsilon +i0^{+}-\epsilon _{s}\left(
k_{x}\right) }.  \label{greens}
\end{align}
\section{Spectrum of the beard edge}\label{app:edge}

For comparison purposes, we also study the edge spectrum for the beard edge. Using a gauge choice where $\mathbf{r}_{g}=({a}/\sqrt{3})\mathbf{e}_{y}$,  and following the same steps as for the zigzag case, we obtain the spectrum for edge state:
\begin{equation}
\epsilon _{s}^{\pm }=\pm \frac{t[2s\lambda _{\mathrm{SO}}\sin \alpha +\cos
\alpha (\lambda _{v}+2s\lambda _{\mathrm{SO}}\sin 2\alpha )]}{\sqrt{(t\cos
\alpha )^{2}+(2\lambda _{\mathrm{SO}}\sin \alpha )^{2}}}.
\end{equation}
In this case, the constraint becomes 
\begin{equation}
-D_{0}=4t^{2}\cos ^{2}\alpha \sin \beta
_{1}\sin \beta _{2},
\end{equation}
where 
\begin{align}
D_{0}&=t^{2}w_{1}w_{2}+u_{1}u_{2}-\epsilon ^{2},\notag \\
 w_{i}&=1+2\cos \alpha \cos \beta _{i},\notag \\
u_{i} &=\lambda _{v}+2s\lambda _{%
\mathrm{SO}}\sin 2\alpha -4s\lambda _{\mathrm{SO}}\sin \alpha \cos \beta _{i}.
\end{align}
The resulting band structure is shown in Fig.~\ref{spectrum-B}.
Note that the edge states intersect at $k_{x}=0$ \cite{Zhang2014Nano}.

\section{Renormalization group analysis}\label{app:rg}
Next, in order to deal with the effects of interactions in a nonperturbative way, we shall rely upon the bosonization technique. The resulting model is analyzed along the lines of the analysis reported is Ref.~\onlinecite{Goldstein2010prlsm}.

 In bosonization the electron field operator for the
right ($R$) and left moving ($L$) edge electron can be expressed in terms of a set of bosonic fields $\theta(x)$ and $\phi(x)$ as follows:
\begin{equation}
\psi _{R(L)}\left( x\right) =\frac{U_{R(L)}}{\sqrt{2\pi v \xi}}
e^{-i\left[ \pm \phi \left( x\right) -\theta \left( x\right) \right] },
\end{equation}
where $\xi$ is a short-distance cutoff, $v$ is the plasmon velocity (cf. Eq.~\ref{eq:plasmon}), $U_{R}$ and $U_{L}$ are the so-called Klein factors satisfying
$\left\{U_{r},U_{r^{\prime}}\right\} =2\delta_{r,r^{\prime}}$, which allows to satisfy the anti-commutation relations between the two fermion chiralities $R$ and $L$.  The bosonic fields obey 
\begin{equation}
\left[ \phi \left( x\right),\theta \left( x^{\prime }\right) \right] =i\frac{\pi}{2} \mathrm{sgn}(x^{\prime}-x).
\end{equation}
 The chiral densities are given by
\begin{equation}
\rho_{R(L)}\left( x\right) =-\frac{1}{2\pi }\left( \partial _{x}\phi \mp
\partial _{x}\theta \right).
\end{equation}

After bosonizing the low energy effective model and upon applying a unitary transformation generated by 
\begin{equation}
S=\exp \left[
i\zeta \theta_0 \right] \end{equation} 
with $\zeta = \delta_F
\left(d^{\dagger }d-\frac{1}{2}\right)$, $\delta_F = \tfrac{KU_{F}}{\pi v}$, and using the factor $e^{-i\zeta \theta \left(
0\right) }\partial _{x}\phi \left( x\right) e^{i\zeta \theta \left( 0\right)
}=\partial _{x}\phi \left( x\right) -i\zeta \left[ \theta \left( 0\right)
,\partial _{x}\phi \left( x\right) \right] =\partial _{x}\phi \left(
x\right) +\zeta \pi \delta \left( x\right) $, the forward scattering term $\propto U_F$ can be
eliminated from $H^{\prime}_{\mathrm{eff}}$ (cf. Eq. \eqref{eq:hamkf}), and the resulting Hamitonian, $H_{\text{eff}}^{\prime \prime
}=S^{\dagger }H_{\text{eff}}^{\prime }S$ reads:
\begin{align}
H_{\text{eff}}^{\prime \prime } &= H_{\ast }+ \frac{v_B}{\xi}
\left[U_{R}U_{L}e^{2i\phi _{0}}+U_{L}U_{R}e^{-2i\phi _{0}}  \right] \notag \\
&+\frac{2y_{B}}{\xi} \left(d^{\dagger }d-
\frac{1}{2}\right)\left[U_{R}U_{L}e^{2i\phi _{0}}
+  U_{L}U_{R}e^{-2i\phi _{0}} \right]\notag \\
&+\frac{y_{t}}{\xi} \left[ 
d^{\dagger}\left(U_{R}e^{-i\left( \phi _{0}-\lambda \theta
_{0}\right) }  -U_{L}e^{i\left( \phi _{0}+\lambda \theta _{0}\right)} \right) \right. \notag\\
&\left. + \quad \left(U_{R}e^{i\left( \phi _{0}-\lambda \theta
_{0}\right) }  -U_{L}e^{-i\left( \phi _{0}+\lambda \theta _{0}\right)} \right)d \right], 
\end{align}
where 
\begin{align}
H_{\ast } &=\frac{v}{2\pi }\int dx\left[ K\left( \partial _{x}\theta
\right) ^{2}+K^{-1}\left( \partial _{x}\phi \right) ^{2}\right]\notag \\ 
& \qquad -\varepsilon
_{0}\left(d^{\dagger }d-\frac{1}{2}\right). 
\end{align}
Here $\varepsilon_0$ denotes the distance of the bound state from the Fermi energy of the edge channel, $\epsilon_F$.   In what follows we focus
on the resonant case for which $\epsilon_0 = 0$.
In addition,  $\lambda =1-\delta_F$, $\phi _{0}=\phi (x=0)$,  $\theta_{0}=\theta (x = 0)$, $v_B$, $y_{B}$, $y_{t}$ are dimensionless couplings,  and $K$ is the Luttinger parameter and $v$ is the edge plasmon velocity.

\begin{figure*}[btp]
\begin{center}
\includegraphics[width=12 cm]{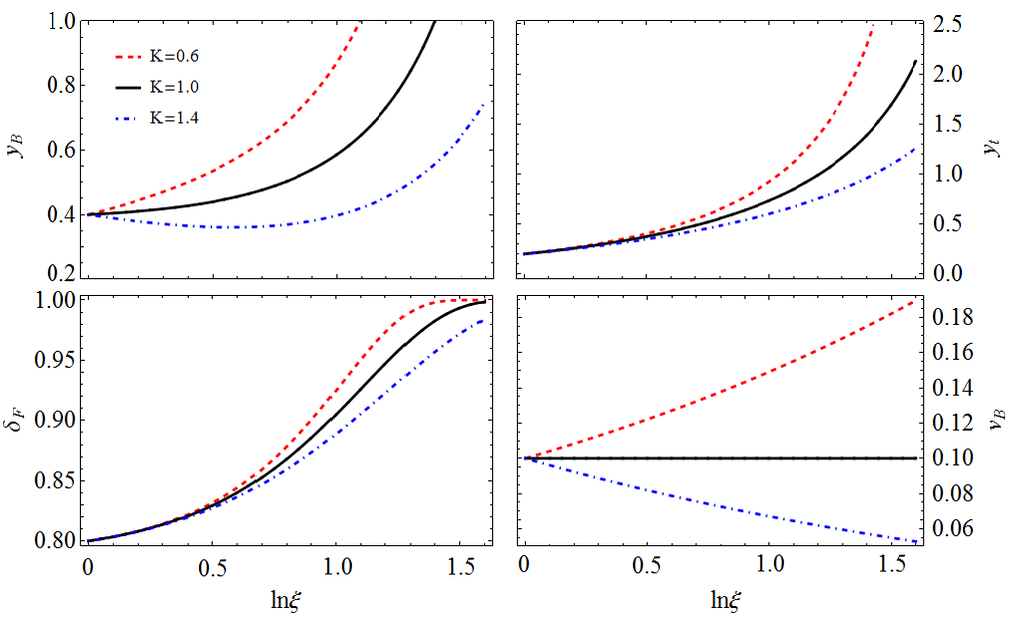} %
\end{center}
\caption{The sketch of the RG flows for the couplings that parametrize the effective low-energy model.}
\label{fig15}
\end{figure*}

 Using Cardy's approach~\cite{Cardy} and taking into account that  
\begin{align}
\left\langle e^{2i\phi_0 (\tau) }e^{-2i\phi_0(0)
}\right\rangle &\sim |\tau| ^{-2K},\\
\left\langle e^{i[ \phi_0 (\tau) -\lambda \theta_0 (\tau) ] }e^{-i[\phi_0 (0) -\lambda \theta_0 (0) ]
}\right\rangle &\sim |\tau| ^{-\alpha(K,\lambda)},\\
\alpha(K,\lambda) = \frac{K}{2}+\frac{\lambda^{2} K^{-1}}{2}, 
\end{align}
we arrive at the set of  RG equations valid to second order in the couplings describing backscattering and tunneling in and out of the resonant level given in ~\eqref{eq:rg1}-\eqref{eq:rg4}.
 The RG equations are similar to those derived in Ref.~\onlinecite{Goldstein2010prlsm} for a model of a resonant level that is side-coupled to an interacting 1D electron system. As described in the main text, the equations show that for weak to moderate attractive interactions (i.e. $K \gtrsim 1$), the tunneling operator $\propto y_t$ is flows to strong coupling. On the other hand, both the backscattering interaction ($\propto y_B$) and potential ($\propto v_B$) will be initially suppressed. Eventually, the runaway flow
of $y_t$ drags along $\delta_F$ and $y_B$, quickly driving the forward interaction with the level to its fixed point $\delta^*_F = 1$. As a result, the transmission through the impurity will be suppressed,
as discussed in the main text. 

  Fig.~\ref{fig15} shows a sketch of the typical RG flows for moderately repulsive (i.e. $K\lesssim 1$) and moderately attractive  (i.e. $K\gtrsim  1$) interactions. In both regimes, alls couplings (execpt for the backscattering potential $v_B$  for $K > 1$, cf Eq.~\ref{eq:rg4}) rapidly reach values of order unity, which in the perturbative approach corresponds to a runaway flow to strong coupling.

\end{document}